\documentclass[journal,12pt, draftclsnofoot, onecolumn]{IEEEtran}%

\IEEEoverridecommandlockouts
\usepackage[reqno]{amsmath}
\usepackage{amssymb}
\usepackage{amsthm}
\usepackage{graphics,graphicx}
\usepackage{fancyhdr}
\usepackage{xcolor}
\usepackage{subfigure}
\usepackage{amsfonts}
\usepackage{graphicx}
\usepackage{bbm}
\usepackage{algorithm}
\usepackage{algorithmic}%
\usepackage{bm}
\usepackage{setspace}
\usepackage{epstopdf}
\usepackage{enumerate}
\usepackage{url}
\providecommand{\U}[1]{\protect\rule{.1in}{.1in}}
\newtheorem{mytheorem}{Theorem}

\theoremstyle{definition}

\newtheorem{myremark}[mytheorem]{Remark}

\title{Kernel-Based Adaptive Online Reconstruction of Coverage Maps
  With Side Information} 
\author{\IEEEauthorblockN{Martin Kasparick\IEEEauthorrefmark{1},
Renato L. G. Cavalcante\IEEEauthorrefmark{1},
Stefan Valentin\IEEEauthorrefmark{2},\\ 
S\l awomir Sta\'nczak\IEEEauthorrefmark{1}, and
Masahiro Yukawa\IEEEauthorrefmark{3}}\\
\IEEEauthorblockA{\IEEEauthorrefmark{1}Fraunhofer Heinrich Hertz Institute, Germany}\\
\IEEEauthorblockA{\IEEEauthorrefmark{2}Bell Labs, Alcatel-Lucent, Germany}\\
\IEEEauthorblockA{\IEEEauthorrefmark{3}Department of Electronics and Electrical Engineering, 
Keio University, Japan}%
  \thanks{ The work was partly supported by Alcatel-Lucent within project PreReAl2,  
    by the European Commission within FP7 project ICT-317669
    METIS, and by the KDDI Foundation.}}

\begin{document}
\onehalfspacing

\maketitle

\begin{abstract}
  In this paper, we address the problem of reconstructing coverage
  maps from path-loss measurements in cellular networks. We propose
  and evaluate two kernel-based adaptive online algorithms as an
  alternative to typical offline methods. The proposed algorithms are
  application-tailored extensions of powerful iterative methods such
  as the adaptive projected subgradient method and a state-of-the-art
  adaptive multikernel method. Assuming that the moving trajectories
  of users are available, it is shown  how side information
  can be incorporated in the algorithms to improve their convergence
  performance and the quality of {the} estimation. The complexity is
  significantly reduced by imposing sparsity-awareness in the sense
  that the algorithms exploit the compressibility of the measurement
  data to reduce the amount of data which is saved and
  processed. Finally, we present extensive simulations based on
  realistic data to show that our algorithms provide fast, robust
  estimates of coverage maps in real-world scenarios. Envisioned
  applications include path-loss prediction along trajectories of
  mobile users as a building block for anticipatory buffering or
  traffic offloading.
\end{abstract}

\section{Introduction}

A reliable and accurate path-loss estimation and prediction, e.g., for the 
reconstruction of coverage maps, have attracted a great deal of
attention from both academia and industry; in fact the problem is
considered as one of the main challenges in the design of radio
networks \cite{SurveyPL13}\cite{SPM14_LocationAware5G}.  A reliable reconstruction of current and
future coverage maps will enable future networks to better utilize
scarce wireless resources and {to} improve the quality-of-service (QoS)
experienced by the users. Reconstructing coverage maps is of
particular importance in the context of (network-assisted)
device-to-device (D2D) communication where no or only partial
measurements are available for D2D channels \cite{Fodor12}. In this
case, some channel state information must be extracted or
reconstructed from other measurements available in the network and
provided to the D2D devices for a reliable decision making
process. Another interesting concept where path-loss prediction is
expected to play a crucial role is proactive resource allocation
\cite{GamalProactive13}\cite{Proebster2012}. Here, the decisions on
resource allocation are based not only on present channel state
information, but also on information about future propagation
conditions. In particular, the quality of service experienced by
mobile users can be significantly enhanced if the information about
future path-loss propagation conditions along the users' routes is
utilized for proactive resource allocation. Possible applications of
proactive resource allocation are:
\begin{enumerate}[i)]
\item Anticipatory buffering: Assure smooth media streaming by filling
  the playout buffer before a mobile user reaches a poorly covered
  area.
\item Anticipatory handover: Avoid handover failures by including a
  coverage map in the decision-making process.
\item Anticipatory traffic offloading: Reduce cellular traffic load by
  postponing transmission of a user until it has reached an area
  covered by WLAN.
\end{enumerate}

In this paper, we consider an arbitrary cellular network, and {we} address
the problem of estimating a two-dimensional map of path-loss
values\footnote{In general, a two-dimensional map of path-loss values
  (also called path-loss map) is a map under which every point in a
  coverage area is assigned a vector of path-losses from this point to
  a subset of base stations.}  at a large time scale. The path-loss
estimation is the key ingredient in the reconstruction of coverage
maps, which provides a basis for the development of {the} proactive resource
allocation schemes mentioned above. We propose novel kernel-based
adaptive online methods by tailoring powerful approaches from machine
learning to the application needs.
In the proposed schemes, each base station maintains a prediction of
the path-loss in its respective area of coverage.  {This} prediction
process is carried out as follows.  We assume that each base station
in the network has access to measurements  {that} are periodically
generated by users in the network. %
Whenever a new measurement arrives, the corresponding base station
updates its current approximation of the unknown path-loss function in
its cell.  To make this update quickly available to online adaptation 
function in the network, coverage map reconstruction needs to be an
online function itself.  Thus, we need to consider machine learning
methods {that} provide \textit{online adaptivity} at \textit{low
  complexity}.  The requirement of low complexity is particularly
important because measurements {arrive} continuously, and they
have to be processed in real time. The online nature of the algorithm
in turn ensures that measurements are taken into account immediately
after their arrivals to continuously improve the prediction and
estimation quality. In addition, we exploit \textit{side information}
to further enhance the quality of the prediction.
{%
By side information, we understand any information in addition
to concrete measurements at specific locations. 
In this paper, we exploit knowledge about the user's trajectories, and we show how 
this information can be incorporated in the algorithm.} 
Measurements used as input to the learning algorithm can be erroneous
due to various causes.
One source of errors results from the fact that wireless users are
usually able to determine their positions only up to a certain
accuracy.  On top of this, the measured values themselves can also be
erroneous.
Another problem is that users performing the measurements are not
uniformly distributed throughout the area of interest and their
positions cannot be controlled by the network. This leads to a
non-uniform random sampling, {and as a results, data availability} may vary significantly among different areas. {To} handle such situations, 
we exploit side information
to perform robust prediction for areas where no or only few
measurements are available.

Our overall objective is to design an adaptive algorithm that, despite
the lack of measurements in some areas, {has} high estimation
accuracy under real-time constraints and practical impairments.
Kernel-based methods are promising tools in this aspect due to 
their 
amenability to adaptive and low-complexity implementation
\cite{WorldOfProjections}. This paper shows how to enhance such methods so that
they can provide accurate coverage maps. In particular, we design
sparsity-aware kernel-based adaptive algorithms that are able to
incorporate side information.

To demonstrate the advantages of the proposed approaches, we present
extensive simulations for two scenarios. First, 
we confine our attention to
an urban scenario closely related to practical situations; the
scenario is based on realistic path-loss data and realistic emulation
of user mobility. {Second, we consider a campus network with publicly available real 
measurement traces at a fairly high spatial resolution. } 

\subsection{Contribution Summary}
In short, the main contributions of our work are:
\begin{enumerate}
\item We demonstrate how kernel-based methods can be used for adaptive
  online learning of path-loss maps based on measurements from users
  in a cellular network.
\item We compare state-of-the-art methods with single and multiple
  kernels with respect to their accuracy and convergence for
  estimating coverage maps.
\item We show how side information can be incorporated in the
  algorithms to improve their accuracy in a given time period. 
\item We show how to enhance the algorithms by using an iterative
  weighting scheme in the underlying cost function.
\end{enumerate}

\subsection{Related Work}
Accurately predicting path-loss in wireless networks is an ongoing
challenge \cite{SurveyPL13}.  Researchers have proposed many different
path-loss models in various types of networks, and there are also
approaches to estimate the path-loss from measurements
\cite{SurveyPL13}.  
{In \cite{phillips2011efficacy}, wireless network measurements are used to evaluate the performance of a large number of empirical path-loss models
proposed over the last decades. The authors demonstrate that even the best models still produce significant errors when compared to actual measurements.}
Regarding measurement-based learning approaches,
so far mainly Support Vector Machines (SVM) \cite{Pred_SVM_ANN11} or
Artificial Neural Networks (ANN)
\cite{PathlossNeuralNetworks01,Pred_SVM_ANN11} have been employed.
{Aside from that, Gaussian processes \cite{SPM14_LocationAware5G} {and kriging based techniques \cite{EstevezKriging}} have recently been successfully used for the estimation of radio maps.}
{In \cite{giannakis_channelGainMaps_2011}, kriging based techniques have been applied to track channel gain maps in a given geographical area. The proposed \emph{kriged Kalman filtering} algorithm  
allows to capture both spatial and temporal correlations.}
However, all these studies use batch schemes; i.e., all (spatial) data needs to be 
available before applying the prediction schemes.  
{Somehow in between learning and modeling is the low-overhead active-measurement based approach of \cite{robinson2008assessment}, which aims at identifying coverage holes 
in wireless networks. The authors use an a-priori model {that} is fitted based on terrain properties and corrections obtained by measurements, and they 
evaluate their method in real-world WIFI networks (\textit{Google WiFi} and the Center for Technology for All Wireless (\textit{TFA}) at Rice University).
In more detail, in a first phase the path-loss characteristics are learned for each terrain/building type and subsequently used to update corresponding parameters in 
the path-loss model. 
A second phase involves iterative refinement based on measurements close to the coverage boundary.}
{Although \cite{robinson2008assessment} uses measurements to refine {iteratively} 
the current path-loss model in an online fashion,} to the best of the 
authors' knowledge, this paper is the first study to propose online
estimation mechanisms for path-loss maps using kernel-based
methods. {Using such methods, we aim at developing algorithms that can be readily deployed in arbitrary scenarios, without the need of initial tailoring 
to the environment.}
In \cite[Section VI]{SurveyPL13} the authors expect ``general machine learning
approaches, and active learning strategies'' to be fruitful; however, 
``applying those methods to the domain of path loss modeling
and coverage mapping is currently unexplored."

In this {study} we rely on methods such as the Adaptive 
Projected Subgradient Method (APSM), which is a recently developed
tool for iteratively minimizing a sequence of convex cost functions
\cite{yamada2005adaptive}. %
It generalizes Polyak's projected subgradient algorithm
\cite{polyak1969} to the case of time-varying functions, and it can be
easily combined with kernel-based tools from machine learning
\cite{yamada2005adaptive}\cite{WorldOfProjections}.  APSM generalizes well-known
algorithms such as the affine projection algorithm (APA) or normalized
least mean squares (NLMS) \cite{sayed2003fundamentals}.  APSM was
successfully applied to a variety of different problems, for example,
channel equalization, diffusion networks, peak-to-average power ratio
(PAPR) reduction, super resolution image recovery, and nonlinear
beamforming \cite{WorldOfProjections}. %

\subsection{Notation}
Let $\mathbb{R}$ and $\mathbb{Z}$ denote the sets of real numbers and integer numbers, respectively.  
We denote vectors by bold face lower case letters and matrices by bold face upper case letters. Thereby
$[\bm{x}]_i$ stands for the $i$th element of vector $\bm{x}$, and $[\bm{A}]_{i,j}$  stands for the element in the $i$th row and $j$th column of matrix $\bm{A}$.
An inner product between two matrices $\bm{A},\bm{B}\in\mathbb{R}^{l\times m}$ is defined by $\langle\bm{A},\bm{B}\rangle:=\mathrm{tr}(\bm{A}^T\bm{B})$ where $(\cdot)^T$ denotes the transpose operation and 
$\mathrm{tr}(\cdot)$ denotes the trace. The Frobenius norm of a matrix $\bm{A}$ is denoted by $\lVert\bm{A}\rVert=\langle\bm{A},\bm{A}\rangle^{\frac{1}{2}}$, which is the norm induced by the 
inner product given above.
By $\mathcal{H}$ we denote a (possibly infinite dimensional) real Hilbert space with an inner product given by $\langle\cdot,\cdot\rangle$ and an 
induced norm $\lVert \cdot\rVert=\langle \cdot,\cdot\rangle^{\frac{1}{2}}$. Note that in this paper we consider different Hilbert spaces (namely 
of matrices, vectors, and functions) 
but for notational convenience we use the same notation for their corresponding inner products and induced norms.
(The respective Hilbert space will be then obvious from the context.) %
A set $C\subset\mathcal{H}$ is called convex if $\lambda \bm{y}_1+(1-\lambda)\bm{y}_2\in C$ $\forall \bm{y}_1,\bm{y}_2\in C,\; \forall\lambda\in(0,1)$.
We use $d(\bm{x},C)$ to denote the Euclidean distance of a point $\bm{x}\in\mathcal{H}$ to a closed convex set $C$, which is given by 
\[
 d(\bm{x},C) := \inf \{\lVert\bm{x}-\bm{y}\rVert:\bm{y}\in C\}.
\]
Note that the infimum is achieved by some $\bm{y}\in C$ {because} we are dealing with closed convex sets in a Hilbert space.

\subsection{Definitions} \label{sec:Definitions}
The projection of $\bm{x}$ onto a nonempty closed convex set $C$, denoted by 
$P_C(\bm{x})\in C$, is the uniquely existing point in the set $\mathrm{argmin}_{\bm{y}\in C}\lVert \bm{x}-\bm{y}\rVert$. 
{Later, if $\mathrm{argmin}_{\bm{y}\in C}\lVert \bm{x}-\bm{y}\rVert$ is a singleton, the same notation is also used to denote the 
unique element of the set.}
A Hilbert space $\mathcal{H}$ is called a reproducing kernel Hilbert space (RKHS) if there exists a (kernel) function $\kappa: \mathbb{R}^m\times\mathbb{R}^m \rightarrow \mathbb{R}$ for which the following properties hold \cite{shawe2004kernel}:
\begin{enumerate}
\item $\forall \bm{x}\in\mathbb{R}^m$, the function $\kappa(\bm{x},\cdot):\mathbb{R}^m\rightarrow\mathbb{R}$ belongs to $\mathcal{H}$, and
\item $\forall\bm{x}\in\mathbb{R}^m$, $\forall f\in\mathcal{H}$, $f(\bm{x})=\langle f,\kappa(\bm{x},\cdot)\rangle$, 
\end{enumerate}
where 2) is called the reproducing property.
Calculations of inner products in a RKHS can be carried out by using the so-called ``kernel trick'':
\[
\langle\kappa(\bm{x}_i,\cdot),\kappa(\bm{x}_j,\cdot)\rangle = \kappa(\bm{x_i},\bm{x_j}).
\]
Thanks to this relation, inner products in the high (or infinite) dimensional feature space can be calculated 
by simple evaluations of the kernel function in the original parameter space. 

The proximity operator $\mathrm{prox}_{\gamma f}:\mathcal{H}\rightarrow\mathcal{H}$ of a scaled function $\gamma f$, with $\gamma>0$ and a continuous %
convex function $f:\mathcal{H}\rightarrow\mathbb{R}$, is defined as 
\[
\mathrm{prox}_{\gamma f}(\bm{x}) = \mathrm{argmin}_{\bm{y}} \left(f(\bm{y}) + \frac{1}{2\gamma}\lVert \bm{x}-\bm{y}\rVert^2\right). 
\]
Proximity operators have a number of properties that are particularly suited for the design of iterative algorithms. In particular,
the proximity operator 
is firmly nonexpansive \cite{Combettes11}, i.e. $(\forall\bm{x},\bm{y}\in\mathcal{H})$
\begin{small}
\begin{equation}
   \lVert\mathrm{prox}_f\bm{x}-\mathrm{prox}_f\bm{y}\rVert^2+\lVert(\bm{x}-\mathrm{prox}_f\bm{x})-(\bm{y}-\mathrm{prox}_f\bm{y})\rVert^2\leq\lVert\bm{x}-\bm{y}\rVert^2   \label{eq:prox_nonexpansive}
\end{equation}
\end{small}
and its fixed point set is the set of minimizers of $f$.

\subsection{Organization} 
The paper is organized as follows. In Section \ref{sec:Problem} we
specify the system model and the path-loss estimation problem.  In
Section \ref{sec:algorithm} we outline important mathematical
concepts and the basics of the employed kernel-based methods. Moreover we present our modifications to the basic
algorithms to incorporate side information and to improve sparsity. In
Section \ref{sec:simulations} we numerically demonstrate the
performance of the algorithms 
{in realistic urban and campus network scenarios.} 

\section{Problem Statement} \label{sec:Problem} 

We now formally state the problem we address in this study. %
{We pose the problem of reconstructing path-loss maps from measurements as a regression problem,  
and {we} use the users' locations as input to the regression {task}.  
This is mainly motivated by the increasing availability of such data in modern wireless networks due to 
the wide-spread availability of low-cost GPS devices. 
Moreover, path-loss is a physical property that is predominantly determined by the locations of transmitters and receivers. 
However, the methods we propose are general enough to be used with arbitrary available input. 
In addition, our methods permit the inclusion of further information as {side information}. We exemplify 
this {fact} in Section \ref{sec:algorithm_distweight}, where we incorporate knowledge about the users' trajectories.} 
 
To avoid technical digressions and notational clutter, we consider that all variables in Sections~\ref{sec:Problem} and \ref{sec:algorithm} are deterministic. Later in Section~\ref{sec:simulations} we drop this assumption to give numerical evidence that the proposed algorithms have good performance also in a statistical sense. \par

We assume that a mobile operator observes a 
sequence $\left\{\left(\tilde{\bm{x}}_n, y_n\right)\right\}_{n\in\mathbb{N}}\subset\mathbb{R}^2\times\mathbb{R}$, 
where $\tilde{\bm{x}}_n:=\bm{x}_n+\bm{\varepsilon}_x\in\mathbb{R}^2$ is an estimate of a coordinate $\bm{x}_n\in\mathbb{R}^2$ of the field, $\bm{\varepsilon}_x\in\mathbb{R}^2$ is an estimation error, and $y_n\in\mathbb{R}$ 
is {a noisy} path-loss {measurement} at coordinate $\bm{x}_n$ with respect to a particular base station (e.g., the base station with strongest received signal), all for the $n$th measurement reported by an user in the system.  The relation between the path-loss measurement $y_n$ and the coordinate $\bm{x}_n\in\mathbb{R}^2$ is given by
\begin{equation}
\label{eq:noisy_function_observation}
y_n:=f({\bm{x}}_n)+\varepsilon_y,
\end{equation}
where $f:\mathbb{R}^2\to\mathbb{R}$ is an unknown function and $\varepsilon_y\in\mathbb{R}$ is an error in the measurement of the path-loss. Measurements $\left(\tilde{\bm{x}}_n, y_n\right)$ arrive sequentially, and they are reported by possibly multiple users in the network. As a result, at any given time, operators have knowledge of only a finite number of terms of the sequence $\left\{\left(\tilde{\bm{x}}_n, y_n\right)\right\}_{n\in\mathbb{N}}$, and this number increases quickly over time. \par 

The objective of the proposed algorithms is to {\it estimate the function $f$ in an online fashion and with low computational complexity}. By online, we mean that the algorithms should keep improving the estimate of the function $f$ as measurements $\left(\tilde{\bm{x}}_n, y_n\right)$ arrive, ideally with updates that can be easily implemented. \par  

In particular,  in this study we investigate two algorithms having the above highly desirable characteristics. The first algorithm is based on the adaptive projected subgradient method for machine learning \cite{WorldOfProjections,Slavakis2008} (Section~\ref{sec:APSM}), and the second algorithm is based on the more recent multikernel learning technique \cite{Masahiro13} (Section~\ref{sec:MK}). The choice of these particular algorithms are motivated by the fact that they are able to cope with large scale problems where the number of measurements arriving to operators is so large that conventional learning techniques cease to be feasible because of memory and computational limitations. Note that, from a practical perspective, typical online learning algorithms need to solve an optimization problem where the number of optimization variables grows together with the number of measurements $\left(\tilde{\bm{x}}_n, y_n\right)$. The algorithms developed in this study operate by keeping only the most relevant data (i.
e., the  
most 
relevant terms of the sequence $\left\{\left(\tilde{\bm{x}}_n, y_n\right)\right\}_{n\in\mathbb{N}}$) in a set called {\it dictionary} and by improving the estimate of the function $f$ by computing simple projections onto closed convex sets, proximal operators, or gradients of smooth functions. We note that the notion of relevance depends on the specific technique used to construct the dictionary, and, for notational convenience, we denote the dictionary index 
set at time $n$ as
\[
\mathcal{I}_n \subseteq \{n,n-1,\ldots,1\}.
\]

The proposed algorithms are able to treat coordinates as continuous variables, but, in digital computers, we are mostly interested in reconstructing path-loss maps at a discrete set of locations called \textit{pixels}. Therefore, to devise practical algorithms for digital computers, we define $\bm{H}\in\mathbb{R}^{X_1\times X_2}$ ($X_1,
X_2\in\mathbb{N}$) to be the \textit{path-loss matrix}, where each
element of the matrix is the path-loss at a given location.  Based on the current estimate of $f$, we can construct an
approximation $\tilde{\bm{H}}$ of the true path-loss matrix, and we note that this estimate has information about the path-loss of future locations where a set of
users of interest (UOI) are expected to visit.

\section{Kernel-Based Methods With Sparsity for Adaptive Online
  Learning Using Side Information} \label{sec:algorithm}

We now turn the attention to the description of the two kernel-based approaches outlined in
Section \ref{sec:Problem}; namely, APSM \cite{WorldOfProjections} and the multikernel approach with adaptive
model selection \cite{Masahiro13}. %

\subsection{The APSM-based algorithm} \label{sec:APSM} 
In this section we give details on the APSM-based algorithm, a well-known method, tailored to the specific application 
of path-loss estimation. In addition, in Section \ref{sec:algorithm_distweight} we propose a novel scheme to weight measurements and corresponding sets, in order to improve the 
performance in our specific application. 

To develop the APSM-based algorithm, a set-theoretic approach, we start by assuming that the function $f$ belongs to a RKHS $\mathcal{H}$. 
As {the $n$'th} measurement $\left(\tilde{\bm{x}}_n, y_n\right)$ becomes available, we construct a closed convex set $S_n\subset\mathcal{H}$ that contains estimates of $f$ that are consistent with the measurement. A desirable characteristic of the set $S_n$ is that it should contain the {\it estimandum} $f$; i.e., $f\in S_n$. 
In particular, in this study we use the hyperslab
\begin{equation}
\label{eq:consistencyConditionAPSM}
S_n := \{h\in\mathcal{H}:\lvert y_n-\langle h,\kappa(\tilde{\bm{x}}_n,\cdot)\rangle\rvert\leq\varepsilon\} = \{h\in\mathcal{H}:\lvert y_n-h(\tilde{\bm{x}}_n)\rvert\leq\varepsilon\}, 
\end{equation}
where $\varepsilon\ge 0$ is a relaxation factor used to take into account noise in measurements, and $\kappa: \mathbb{R}^2\times\mathbb{R}^2\to\mathbb{R}$ is the kernel of the RKHS $\mathcal{H}$. In the following, we assume that $f\in S_n$ for every $n\in\mathbb{N}$. 

Unfortunately, a single set $S_n\subset\mathcal{H}$ is unlikely to contain enough information to provide reliable estimates of the function $f$. More precisely, the set $S_n\subset\mathcal{H}$ typically contains vectors $\tilde{f}\in S_n$ that are far from $f$, in the sense that the distance $\|\tilde{f}-f\|$ is not sufficiently small to consider $\tilde{f}$ as a good approximation of $f$. However, we can expect an arbitrary point in the closed convex set $S^\star:=\cap_{n\in\mathbb{N}} S_n\ni f$ to be a reasonable estimate of $f$ because $S^\star$ is the set of vectors that are consistent with every measurement we can obtain from the network. As a result, in this set-theoretic approach, we should aim at solving the following convex feasibility problem: 
\\[.5cm]
{\it Find a vector $\tilde{f}\in\mathcal{H}$ satisfying $\tilde{f}\in\cap_{n\in\mathbb{N}} S_n$.}
\\[.5cm]

In general, solving this feasibility problem is not possible because, for example, we are not able to observe and store the whole sequence $\left\{\left(\tilde{\bm{x}}_n, y_n\right)\right\}_{n\in\mathbb{N}}$ in practice (recall that, at any given time, we are only able observe a finite number of terms of this sequence). As a result, the algorithms we investigate here have the more humble objective of finding an arbitrary vector in the set $S:=\overline{\bigcup_{t=0}^{\infty}\bigcap_{n>t} C_n} \ni f$, where, at time $n$, $C_n$ is the intersection of selected sets from the collection $\{S_1,\dots, S_n\}$ (soon we come back to this point). We can also expect such a vector to be a reasonable estimate of $f$ because $S$ corresponds to vectors that are consistent with all but finitely many measurements. \par 

 Construction of the set $S$ is also not possible because, for example, it uses infinitely many sets $C_n$. However, under mild assumptions, algorithms based on {the} APSM are able to produce a sequence $\{\hat{f}_n\}_{n\in\mathbb{N}}\subset\mathcal{H}$ of estimates of $f$ that i) can be computed in practice, ii) converges asymptotically to an unspecified point in $S$, and iii) has the monotone approximation property (i.e., $\|\hat{f}_{n+1}-f\|<\|\hat{f}_{n}-f\|$ for every $n\in\mathbb{N}$). \par 

 In particular, in this study we propose a variation of the adaptive projected subgradient method described in 
\cite{WorldOfProjections,Slavakis2008}. In more detail, at each iteration $n$, we select $q$ sets from the collection $\{S_1,\ldots,S_n\}$ with the approach described in \cite{WorldOfProjections}. The intersection of these sets is the set $C_n$ described above, and the index of the sets chosen from the collection is denoted by 
\begin{equation}
\mathcal{I}_{n,q} := \{i^{(n)}_{r_n},i^{(n)}_{r_n-1},\ldots,i^{(n)}_{r_n-q+1}\} \subseteq \{1,\ldots,n\}, \label{eq:APSM_indexset}
\end{equation}
where $n\geq q$ and $r_n$ is the size of dictionary. 
{(Recall that the dictionary is simply the collection of ``useful'' measurements, which 
are stored to be used later in the prediction. Thus, in our case, the dictionary comprises a set of GPS coordinates corresponding to  the respective path-loss measurements.)} With this selection of sets, starting from $\hat{f}_0=0$, 
we generate a sequence $\{\hat{f}_{n}\}_{n\in\mathbb{N}}\subset\mathcal{H}$ by 
\begin{equation}
\hat{f}_{n+1} := \hat{f}_n + \mu_n\left(\sum_{j\in\mathcal{I}_{n,q}}w_{j,n}P_{S_j}(\hat{f}_n)-\hat{f}_n\right), \label{eq:UpdateAPSM}
\end{equation}
where $\mu_n\in(0,2\mathcal{M}_n)$ is the step size, $\mathcal{M}_n$ is a scalar given by 
\begin{equation*}
\mathcal{M}_n:=\begin{cases}
		\mbox{\footnotesize $\frac{{\displaystyle\sum_{j\in\mathcal{I}_{n,q}}w_{j,n}\lVert P_{S_j}(f_n)-f_n\rVert^2}}{{\displaystyle\lVert \sum_{j\in\mathcal{I}_{n,q}} w_{j,n} P_{S_j}(f_n)-f_n\rVert^2}}$}, 
		&\mbox{if \footnotesize $f_n\notin \bigcap_{j\in\mathcal{I}_{n,q}} S_j$},\\   %
		1, &\text{otherwise,}
		\end{cases} %
\end{equation*}
and $w_{j,n}>0$ are weights satisfying 
\begin{equation}
\sum_j w_{j,n}=1. \label{eq:WeightConditionAPSM}
\end{equation}
The projection onto the hyperslab induced by measurement $n$ is given
by $P_{S_n}(f)=f+\beta_f \kappa(\tilde{\bm{x}}_n,\cdot)$ where
\[
\beta_f:= \begin{cases}
	  \frac{y-\langle f,\kappa(\tilde{\bm{x}}_n,\cdot)\rangle-\varepsilon}{\kappa(\tilde{\bm{x}}_n,\tilde{\bm{x}}_n)},&\text{if } \langle f,\kappa(\tilde{\bm{x}}_n,\cdot)\rangle-y<-\varepsilon,\\
	  \frac{y-\langle f,\kappa(\tilde{\bm{x}}_n,\cdot)\rangle+\varepsilon}{\kappa(\tilde{\bm{x}}_n,\tilde{\bm{x}}_n)},&\text{if } \langle f,\kappa(\tilde{\bm{x}}_n,\cdot)\rangle-y>\varepsilon,\\
	  0,&\text{if } \lvert\langle f,\kappa(\tilde{\bm{x}}_n,\cdot)\rangle-y\rvert\leq\varepsilon.\\
	  \end{cases}
\]

For details of the algorithm, including its geometrical interpretation 
we refer the reader to \cite{WorldOfProjections}. For the sparsification of the dictionary we use the heuristic %
described in \cite{Slavakis2008}. Here, we focus on the choice of weights, which can be efficiently exploited in the proposed application {domain} to improve the performance of the algorithm.

\subsection{Weighting of parallel projections based on side information}\label{sec:algorithm_distweight}

Assigning a large weight $w_{j,n}$ (in comparison to the remaining weights) to the projection $P_{S_j}$ in (\ref{eq:UpdateAPSM}) has the practical consequence that the update in (\ref{eq:UpdateAPSM}) moves close to the set $S_j$. Therefore, previous studies recommend to give large weights to reliable sets. However, in many applications, defining precisely what is meant by reliability is difficult, so uniform weights $w_{j,n}=1/q$ are a common choice \cite{Slavakis2008}. In the proposed application, although we do not define a notion of reliability, we can clearly indicate which sets are the most important for the updates. For instance, sets corresponding to measurements taken at pixels farther away from the route of the UOI should be given smaller weights than measurements of pixels that are close to the user's trajectory. The reason is that estimates should be accurate at the pixels the UOI is expected to visit because these are the pixels of interest to most applications (e.g., video caching based on 
channel conditions). 
Therefore, we assign large weights to measurements close to the UOI's route by proceeding as follows. 
 Let $\mathcal{X}_{\mathrm{UOI}}\subset\mathbb{N}^2$ be the set of pixels that belong to the path of the UOI. Then, for each weight $w_{i,n}$, we compute 
\begin{equation}
w_{i,n} = \frac{1}{d_{\mathrm{min}}(\tilde{\bm{x}}_i, \mathcal{X}_{\mathrm{UOI}})+\varepsilon_w}, \label{eq:WeightsSideInfoAPSM}
\end{equation}
where $d_{\mathrm{min}}(\tilde{\bm{x}}_i, \mathcal{X}_{UOI})$ denotes the minimum distance of measurement $\tilde{\bm{x}}_i$ to the area of interest, and $\varepsilon_w>0$ is  
a small regularization parameter. This distance can be obtained for each pixel $\tilde{\bm{x}}_i$  by considering the distances of every pixel in $\mathcal{X}_{\mathrm{UOI}}$
to $\tilde{\bm{x}}_i$ and by taking the minimum of these distances.
Subsequently, the weights are normalized to ensure the condition shown in (\ref{eq:WeightConditionAPSM}).
To improve the performance in cases with varying data or to exclude pixels that the user has already been present, we can also replace the set $\mathcal{X}_{\mathrm{UOI}}$ 
in (\ref{eq:WeightsSideInfoAPSM}) by the area of interest as discussed in Section \ref{sec:Problem}. Compared to an equal choice of the weights, the proposed method provides fast convergence to a given prediction quality for the UOI, but at the cost of degraded performance in other areas of the map.

\subsection{Multi-Kernel Approach} \label{sec:MK}
{We now turn our attention to an alternative approach based on}
 a state-of-the-art multikernel algorithm. {In} Section \ref{sec:algorithm_sparsity} we provide a novel analytical justification 
for an iterative weighting scheme, which previously has been mainly used as a heuristic. %

In the proposed algorithm of Section \ref{sec:APSM}, based on APSM, the choice of the kernel $\kappa$ is left open, but we note that different choices lead to algorithms with different estimation properties. Choosing an appropriate kernel for a given estimation task is one of the main challenges for the 
application of kernel methods{. To} address this challenge in the path-loss estimation problem, we propose the application of the multikernel algorithm described in \cite{Masahiro13}. Briefly, this algorithm provides good estimates by selecting, automatically, both a reasonable kernel (the weighted sum of a few given kernels) and a sparse dictionary.

In more detail, let $\kappa_m$ be a given kernel function from a set indexed by $m\in\mathcal{M}:=\{1,\ldots,M\}$.
At time $n$, the approach assumes that the  path-loss function $f$ can be approximated by
\begin{equation}
\hat{f}_n (\bm{x}) = \sum_{m\in\mathcal{M}}\sum_{i=1}^{|\mathcal{I}_n|}\alpha_{i,n}^{(m)}\kappa_m(\tilde{\bm{x}}_i,\bm{x}) \label{eq:approx_Multikernel}
\end{equation}
if $\alpha_{i,n}^{(m)}\in\mathbb{R}$ are appropriately chosen scalars. %

At coordinate $\bm{x}_n$,  (\ref{eq:approx_Multikernel}) can be equivalently written as 
$\hat{f}_n (\bm{x}_n) = \langle\bm{A}_n,\bm{K}_n\rangle$, where $n$ is the time index, $r_n = |\mathcal{I}_n|$ is the size of the dictionary $\mathcal{I}_n$, 
and $\bm{A}_n$ and $\bm{K}_n\in \mathbb{R}^{M\times r_n}$ are matrices given by 
$[\bm{A}_n]_{m,i}:=\alpha_{j_i^{(n)},n}^{(m)}$ and $[\bm{K}_n]_{m,i}=\kappa_m\left(\bm{x}_n, \tilde{\bm{x}}_{j_i^{(n)}}\right)$, respectively.
Here, $j_i^{(n)}$ denotes the element that at time $n$ is at the $i$th position of the dictionary. In addition, matrices $\tilde{\bm{A}}_n$, $\tilde{\bm{K}}_n\in \mathbb{R}^{M\times r_{n+1}}$
incorporate the update of the dictionary set {(the actual update procedure is described at the end of this section)}, 
thus, $[\tilde{\bm{A}}_n]_{m,i}:=\alpha_{j_i^{(n+1)},n}^{(m)}$ and $[\tilde{\bm{K}}_n]_{m,i}=\kappa_m\left(\bm{x}_n, \tilde{\bm{x}}_{j_i^{(n+1)}}\right)$.
Let us further define 
$d(\bm{A},S_n):=\min_{\bm{Y}\in S_n}\lVert\bm{A}-\bm{Y}\rVert$,
where 
\begin{equation}
S_n:= \left\{\bm{A}\in\mathbb{R}^{M\times r_{n+1}}:\lvert\langle\bm{A},\tilde{\bm{K}}_n\rangle-y_n\rvert\leq\varepsilon_{\mathrm{MK}}\right\} \label{eq:HyperslabMK}
\end{equation}
is a hyperplane defined for matrices, and $\varepsilon_{\mathrm{MK}}$ is a relaxation factor to take into account noise in measurements. For notational convenience, we denote the $i$th column of $\bm{A}$ as $\bm{a}_i$ and the
$m$th row of $\bm{A}$ as $\bm{\xi}_m$. At time $n$, similarly to the study in \cite{Masahiro13}, the coefficients $\alpha_{i,n}^{(m)}\in\mathbb{R}$ are obtained by trying to minimize 
the following cost function:
\begin{equation}
  \Theta_n(\bm{A}):= \underbrace{\frac{1}{2}d^2(\bm{A},S_n)}_{\phi_n(\bm{A})} + \underbrace{\lambda_1 \sum_{i=1}^{r_{n+1}}w_{i,n}\lVert \bm{a}_i\rVert}_{\psi_n^{(1)}(\bm{A})}  + \underbrace{\lambda_2 \sum_{m=1}^{M}\nu_{m,n}\lVert \bm{\xi}_m\rVert}_{\psi_n^{(2)}(\bm{A})}, \label{eq:CF-MK}
\end{equation}
where $\lambda_1$ and $\lambda_2$ are positive scalars used to trade how well the model fits the data, the dictionary size, and the number of atomic kernels $\kappa_m$ being used. In turn, $w_{i,n}$ and $\nu_{m,n}$ are positive weights that can be used to improve sparsity in the dictionary and in the choice of kernels, respectively. Note that the first term $\phi_n(\bm{A})$ is responsible for fitting the function to the
training set. The second term $\psi_n^{(1)}(\bm{A})$ is used to discard
irrelevant data points over time, thus promoting sparsity in the
dictionary (even when the underlying data is changing). In turn, the third term $\psi_n^{(2)}(\bm{A})$ is designed to reduce the influence of
unsuitable kernels.  This provides us with an online model selection
feature that not only provides a high degree of adaptivity, but it also
helps to alleviate the overfitting problem \cite{Masahiro13}.

The main challenge in minimizing (\ref{eq:CF-MK}) is that the optimization problem changes with each new measurement (note the presence of the index $n$ in the cost function). As a result, we cannot hope to solve the optimization problem with simple iterative schemes at each $n$ because, whenever we come close to a solution
of a particular instance of the optimization problem, it may have already changed because new measurements are already available.
However, we hope to be able to track solutions of the time-varying optimization problem for $n$ sufficiently large by following the reasoning of the forward-backward splitting method 
{at each update time}.

In more detail, consider the time-invariant optimization problem: 
\begin{equation}
 {\min.}_{\bm{x}\in\mathcal{H}} \phi(\bm{x}) + \psi(\bm{x}), \label{eq:CF_ProxForwBack}
\end{equation}
where $\phi$, $\psi$ are lower semicontinuous convex functions, where $\phi$ is a differentiable function and $\psi$ is possibly non-differentiable.
We also assume that the set of minimizers is nonempty and $\nabla\phi$ is Lipschitz continuous with Lipschitz constant $L$.
By using properties of the proximal operator described in Section \ref{sec:Definitions}, the following iterative algorithm can converge to a 
solution of (\ref{eq:CF_ProxForwBack}) \cite{yamada2011minimizing}
\begin{equation}
 \bm{x}_{n+1} := \mathrm{prox}_{\frac{\mu}{L}\psi}\left(\bm{I} - \frac{\mu}{L}\nabla\phi\right)(\bm{x}_n), \label{eq:iter_ProxForwBack}
\end{equation}
with step size $\mu\in(0,2)$. For fixed problems such as that in (\ref{eq:CF_ProxForwBack}),  the sequence $\{\bm{x}_{n+1}\}$ produced by (\ref{eq:iter_ProxForwBack}) converges to the solution of (\ref{eq:CF_ProxForwBack}). To see this, note that, by using Property (\ref{eq:prox_nonexpansive}), the operator $T:=\mathrm{prox}_{\frac{\mu}{L}\psi}\left(\bm{I} - \frac{\mu}{L}\nabla\phi\right)$
is a concatenation of two $\alpha$-averaged nonexpansive\footnote{{Recall that a} firmly nonexpansive mapping 
is an $\alpha$-averaged nonexpansive mapping with $\alpha=\frac{1}{2}$ \cite{yamada2011minimizing}.
} operators $T_1:= \mathrm{prox}_{\frac{\mu}{L}\psi}$ and $T_2:=\bm{I} - \frac{\mu}{L}\nabla\phi$, i.e., $T=T_1T_2$.
Convergence of (\ref{eq:iter_ProxForwBack}) to a fix point of $T$, which is the solution to (\ref{eq:CF_ProxForwBack}), then follows from \cite[Proposition 17.10]{yamada2011minimizing}.

The main idea of the multikernel learning approach is to use the above iteration for fixed optimization problems in adaptive settings, with the hope of obtaining good tracking and estimation capabilities. In our original problem with time varying functions, (\ref{eq:CF-MK}) comprises of a differentiable function with Lipschitz continuous gradient and two non-differentiable
functions for which the proximal operator can be computed easily. In order to apply the proximal forward-backward splitting method outlined above, we first modify $\Theta_n$ to take a similar form to the cost function in (\ref{eq:CF_ProxForwBack}). To this end, we approximate (\ref{eq:CF-MK}) by 
\begin{equation}
  \tilde{\Theta}_n(\bm{A}):=\underbrace{\phi_n(\bm{A})+^\gamma\psi_n^{(1)}(\bm{A})}_{\text{smooth}}+\underbrace{\psi_n^{(2)}(\bm{A})}_{\text{proximable}} \label{eq:CF-MK_Moreau_Approx}
\end{equation}
 with $^\gamma\psi_n^{(1)}(\bm{A})$ being the Moreau envelope of $\psi_n^{(1)}(\bm{A})$ of index $\gamma\in(0,\infty)$ \cite{Masahiro13}.  
The problem of minimizing the function in (\ref{eq:CF-MK_Moreau_Approx}), comprising a smooth and a proximable part,
has now a similar structure to the problem in (\ref{eq:CF_ProxForwBack}). Therefore the following iterative algorithm can be 
derived by using known properties of proximal operators:
\begin{equation}
\bm{A}_{n+1} = \mathrm{prox}_{\eta\psi_n^{(2)}}\left[\tilde{\bm{A}}_n-\eta\left(\nabla\phi_n(\tilde{\bm{A}}_n)+\nabla^\gamma\psi_n^{(1)}(\tilde{\bm{A}}_n)\right)\right]. \label{eq:UpdateMK}
\end{equation}
The step size $\eta\in(0,2/L_2)$ is based on the Lipschitz constant $L_2$ of the mapping $T: \mathbb{R}^{M\times r_{n+1}}\rightarrow\mathbb{R}^{M\times r_{n+1}}$, 
$\bm{A}\mapsto\nabla\phi_n({\bm{A}}_n)+\nabla^\gamma\psi_n^{(1)}({\bm{A}}_n)$.
Note that (\ref{eq:UpdateMK}) is the iteration in (\ref{eq:iter_ProxForwBack}) with time-varying functions. 

The applied sparsification scheme is a combination of two parts (which combine the two approaches from \cite{6203609}). First, a new measurement is added to the dictionary only if it is sufficiently new (similar to the 
sparsification scheme used in the APSM algorithm). {Second, columns of $\bm{A}$ with a Euclidean norm close to zero can be simply removed (we use a threshold of $10^{-2}$ in the numerical evaluations of Section \ref{sec:simulations}). By doing so, irrelevant data is discarded from the dictionary.
For further details on the algorithm and the sparsification scheme, {we} refer {the reader} to \cite{Masahiro13}}.

\subsection{Multi-Kernel: Sparsity based on iterative weighting} \label{sec:algorithm_sparsity}
To improve the performance of the multikernel algorithm, we propose a different weighting of rows and columns of $\bm{A}$ in the sparsity-enforcing parts of the 
cost function (\ref{eq:CF-MK}). 
We employ the idea of iterative re-weighting that has also been used in compressive sensing \cite{Candes2008}.
As a first step to determine the weights of the sparsity related cost-term in (\ref{eq:CF-MK}), we use 
\begin{equation}
\hat{w}_{i,n} = \frac{1}{\lVert \bm{a}_{i,n} \rVert+\varepsilon_{1}},  \label{eq:NewIterativeWeights}
\end{equation}
with $\varepsilon_{1}>0$ being a small regularization parameter to ensure stability.
The weights have to be normalized in order to keep the balance between the different terms in (\ref{eq:CF-MK}), such that the final weights are given by
\[
 w_{i,n} = \frac{\hat{w}_{i,n}}{\bar{w}_n},
\]
where $\bar{w}_n = \sum_i \hat{w}_{i,n}$.
The same iterative weighting can also be applied to the weights $\nu_{m,n}$ of the row sparsity enforcing term of the cost function (\ref{eq:CF-MK}). 

The reasoning behind this approach, which is inspired by majorization minimization algorithms \cite{PCS13eeen}, becomes clear when observing the connection 
to the $l_0$-norm.
Let us go back to the cost function we aim to minimize, which is 
\begin{equation}
 \Theta_n(\bm{A}):= \frac{1}{2}d^2(\bm{A},S_n) +  \lambda_1 \sum_{i=1}^{r_{n+1}}w_{i,n}\lVert \bm{a}_i\rVert. \label{eq:CF-MinMaj}
\end{equation}
For simplicity, we consider only the term inducing column sparsity. Note that the weighted block-$l_1$ norm in (\ref{eq:CF-MinMaj})
to enforce block-sparsity is only used as a simple convex approximation of the $l_0$ (quasi-)norm. 
Therefore, (\ref{eq:CF-MinMaj}) should ideally be replaced by 
\begin{equation}
 \Theta_n(\bm{A}):= \frac{1}{2}d^2(\bm{A},S_n) +  \lambda_1 \lVert\lvert\bm{A}\rvert_\ast^T\bm{1}\rVert_0, \label{eq:CF-MinMa_l0}
\end{equation}
where the $|\cdot|_\ast$ operator stands for element-wise absolute value, and $\bm{1}$ denotes a vector of ones of appropriate dimension.
Using a similar expression for the $l_0$ norm as in  \cite{sriperumbudur11}\cite{PCS13eeen}, 
we can write 
\begin{equation}
  \lVert \lvert\bm{A}\rvert_\ast^T\bm{1} \rVert_{0} =\lim_{\varepsilon_2\rightarrow 0}\sum_i \frac{\log(1+ \lVert \bm{a}_i\rVert\varepsilon_2^{-1})}{\log(1+\varepsilon_2^{-1})}. \label{eq:Log_BlockL0Norm}
\end{equation}
Fixing $\varepsilon_2>0$ in (\ref{eq:Log_BlockL0Norm}), we can obtain the following approximation to the minimization of (\ref{eq:CF-MinMa_l0})
\begin{equation}
  \min_{\bm{A}}. \Biggl[\underbrace{\frac{1}{2}d^2(\bm{A},S_n)  + \lambda {\sum_i \log\left(\varepsilon_2+\lVert\bm{a}_i\rVert\right)}}_{g_0(\bm{A})}\Biggr]. \label{eq:CF-MinMaj_Approx_org}
  \end{equation}
  The constant $\lambda$ incorporates both $\lambda_1$ from
  (\ref{eq:CF-MinMaj}), and the omitted constants from
  (\ref{eq:Log_BlockL0Norm}).  Introducing {a vector of} auxiliary variables 
  $\bm{z}$, {where each element corresponds to a column of $\bm{A}$,} Problem (\ref{eq:CF-MinMaj_Approx_org}) can be
  equivalently written as
\begin{align}
  \min_{\bm{A},\bm{z}}. \quad &\Biggl[\underbrace{\frac{1}{2}d^2(\bm{A},S_n)  + \lambda {\sum_i \log\left(\varepsilon_2+{z_i}\right)}}_{g(\bm{A},\bm{z})}\Biggr],  \label{eq:CF-MinMaj_Approx} \\
  \mathrm{s.t.} \quad & \lVert \bm{a}_{i}\rVert \leq {z}_{i} \quad (\forall i). \nonumber
  \end{align}
  Note that in (\ref{eq:CF-MinMaj_Approx}) we are minimizing the sum
  of a convex and a concave function.

  To address this intractable optimization problem, we use a minimization-majorization
  algorithm, which relies on constructing a convex majorizing
  function.  In more detail, for the general problem
\begin{equation}
\min_{\bm{v}_1,\bm{v}_2}. g_1(\bm{v}_1) + g_2(\bm{v}_2) \text{ s.t. } \bm{v}_1\in\mathcal{C}_1,\bm{v}_2\in\mathcal{C}_2, \label{eq:MinSumConvConc}
\end{equation}
where $\mathcal{C}_1,\mathcal{C}_2$ are convex sets, $g_1$ is a convex function, and $g_2$ a concave function, 
the following iteration can be applied to approximate a solution of (\ref{eq:MinSumConvConc})
\begin{align}
    \left(\bm{v}_1^{(l+1)},\bm{v}_2^{(l+1)}\right)\in&\arg\min_{\bm{v}_1,\bm{v}_2} \hat{g}(\bm{v}_1,\bm{v}_2,\bm{v}_2^{(l)}), \label{eq:MM_iteration} \\ 
  &\text{s.t. } \bm{v}_1\in\mathcal{C}_1,\bm{v}_2\in\mathcal{C}_2,  \nonumber
  \end{align}
 where
 \[
  \hat{g}(\bm{v}_1,\bm{v}_2,\bm{w}) := g_1(\bm{v}_1) + g_2(\bm{w}) + \nabla g_2(\bm{w})^T(\bm{v}_2-\bm{w})
 \]
is a convex majorizer of the function $g(\bm{v}_1,\bm{v}_2):=g_1(\bm{v}_1)+g_2(\bm{v}_2)$ in (\ref{eq:MinSumConvConc}).
This function fulfills the properties
\begin{align*}
 g(\bm{v}_1,\bm{v}_2) &\leq \hat{g}(\bm{v}_1,\bm{v}_2,\bm{w}), \quad \forall \bm{v}_1\in\mathcal{C}_1,\bm{v}_2,\bm{w} \in \mathcal{C}_2 \\
 g(\bm{v}_1,\bm{v}_2)  &=  \hat{g}(\bm{v}_1,\bm{v}_2,\bm{v}_2), \quad \forall \bm{v}_1\in\mathcal{C}_1,\bm{v}_2 \in \mathcal{C}_2
\end{align*}
which can be used to show {that} the iteration in (\ref{eq:MM_iteration}) satisfies $g_1(\bm{v}_1^{(l+1)})+g_2(\bm{v}_2^{(l+1)}) \le g_1(\bm{v}_1^{(l)})+g_2(\bm{v}_2^{(l)})$.

Coming back to our original problem (\ref{eq:CF-MinMaj_Approx}), with
$\left[\nabla g_2(\bm{z})\right]_i =
\frac{\lambda}{\varepsilon+{z_i}}$ and eliminating additive constants,
we have {that the iteration in \eqref{eq:MM_iteration} takes the particular form} 
\begin{align*}
\left(\bm{A}^{(l+1)},\bm{z}^{(l+1)}\right)\in \arg\min_{\bm{A},\bm{z}} &\left[\frac{1}{2}d^2(\bm{A},S_n)  + \lambda \sum_i \frac{z_i }{\varepsilon_2+{z^{(l)}_i}}\right], \\
\mathrm{s.t.} \quad & \lVert \bm{a}_{i}\rVert \leq {z}_{i} \quad (\forall i). 
\end{align*}
Substituting back the auxiliary variable $\bm{z}$, it becomes clear
that we find an approximation to the solution of (\ref{eq:CF-MinMaj_Approx_org}) by using the iteration
 \[
  \bm{A}^{(l+1)} \in  \arg\min_{\bm{A}}\left[\frac{1}{2}d^2(\bm{A},S_n) +  \lambda\sum_{i=1}^{r_n} \frac{\lVert\bm{a}_i\rVert}{\lVert\bm{a}_i^{(l)}\rVert+\varepsilon_2},\right]
 \]
  which produces a monotone nonincreasing sequence $\{g_0(\bm{A}^{(l)})\}$, and the choice of the weights
 (\ref{eq:NewIterativeWeights}) applied to (\ref{eq:CF-MinMaj}) as a means of trying to solve (\ref{eq:CF-MinMa_l0}) becomes apparent.%
  {
  \begin{myremark}
  Although widely used in the literature, the concave approximation in (\ref{eq:CF-MinMaj_Approx_org}) is only one of many options to approximate the $l_0$ norm in (\ref{eq:Log_BlockL0Norm}). 
  For instance, \cite{mangasarian1996machine,rinaldi2011concave} propose algorithms based on various alternative concave approximations. 
  A {formal} comparison between these these approximations is a topic that needs further  study \cite[Sec. 5.2]{Candes2008}, {and this investigation}
  is out of the scope of this paper. 
The chosen approach of iterative reweighting the block-$l_1$ norm is a practical and popular approach to support %
sparsity by eliminating elements that are close to zero. For the sake of completeness, it should be noted that different approaches {have been} suggested in the literature.
For example, the FOCUSS algorithm of \cite{gorodnitsky1997sparse} applies a reweighted $l_2$ minimization at each iteration to enforce sparsity.
However, as was pointed out in \cite{Candes2008}, numerical experiments suggest that the reweighted $l_1$-minimization can recover sparse signals with lower error or fewer measurements.
The author of \cite{rinaldi2011concave} reports  good performance {by} using a concave approximation of the $l_0$ norm together with a Frank-Wolfe type algorithm. 
{In fact, the minimization-majorization algorithm used in the context of iterative reweighting (and
discussed above) is closely related to the Frank-Wolfe algorithm, which at each iteration constructs a linear approximation of the 
objective function at the current position and moves towards a minimizer of this function over the same domain.}
\end{myremark}
}
 
 \subsection{Computational Complexity}
In general, the computational complexity of the APSM method, per
projection, is linear w.r.t. the dictionary size \cite{WorldOfProjections}. {In} addition, 
the correlation-based sparsification scheme has quadratic complexity in the dictionary size
\cite{Slavakis2008}. 
For the
multikernel scheme, as derived in \cite{6203609}, the complexity
regarding memory is given by $(L+M)r_n$, where $L$ denotes the
dimensions of the input space, $M$ is the number of kernels, and $r_n$
is as usual the size of the dictionary at time $n$. Similarly, the computational complexity increases roughly
with a factor of $M$ \cite{6203609}.

\section{Numerical Evaluation} 
\label{sec:simulations}

In this section, we numerically evaluate the two proposed iterative
algorithms by applying them to the problem of estimating path-loss
matrices. 

\subsection{Preliminaries}
\label{sec:simulations_prel}

As described in Section \ref{sec:Problem}, the unknown path-loss
matrix $\bm{H}\in\mathbb{R}^{X_1\times X_2}$ is composed of the values
of the function $f$ evaluated at a given discrete set of
locations. The objective is to estimate $\bm{H}$ from noisy
measurements related to $f$ according to
(\ref{eq:noisy_function_observation}). For simplicity, we assume in
this section that the measurements are i.i.d. uniformly distributed
over the the area and arrive according to a Poisson distribution, with
parameter $\lambda=0.1$. Due to noisy observations, at every iteration
$i\in\mathbb{N}$, an estimate $\tilde{\bm{H}}(i)$ of $\bm{H}$ is a
matrix-valued random variable and we use a normalized mean-square
error (MSE) to measure the quality of the estimation. To be precise,
given a path-loss matrix $\bm{H}\neq 0$, the MSE at iteration $i$ is
defined to be
\begin{equation}
{\mathrm{MSE}_i} := E\Bigl[\frac{1}{\lVert \bm{H}\rVert^2}
\lVert\bm{H} -\tilde{\bm{H}}(i)\rVert^2\Bigr]\,. \label{eq:definition_MSE}
\end{equation}
Since the distribution of $\tilde{\bm{H}}(i)$ is not known, we
approximate $\mathrm{MSE}_i$ by {averaging} a sufficiently large number 
of simulation runs.  As a result, the MSE values
presented in this section are {empirical} approximations of the MSE
defined above.

We consider two communications scenarios with different path-loss
models.
\begin{itemize}
\item \textit{Urban scenario}: A real-world urban network based on a
  pixel-based mobility model and realistic collection of data for the
  city of Berlin \cite{momentum}.
\item {\textit{Campus network scenario}: A campus network with real-world measurements and high spatial resolution.}
\end{itemize}
{While the following subsection is devoted to the urban scenario, the outcome of numerical experiments for the campus scenario are presented
in Subsection \ref{sec:simulations:small_scale}.}
Specific simulation parameters can be found in 
the corresponding subsection, except for kernel functions that are
common for both scenarios. Although our results are applicable to
arbitrary kernels, we confine our attention 
to Gaussian kernels of the form
\begin{equation}
\kappa(\bm{x}_1,\bm{x}_2) := \exp\left(-\frac{\lVert\bm{x}_1-\bm{x}_2\rVert^2}{2\sigma^2}\right)\,, \label{eq:Gaussian}
\end{equation}
where the kernel width $\sigma^2$ is a free parameter. The value of
this and other parameters for the APSM algorithm and for our
multikernel algorithm can be found in Table \ref{tab:param_APSM} and
Table \ref{tab:param_MultiKernel}, respectively.
\begin{table} 
\caption{APSM Simulation Parameters}%
\label{tab:param_APSM}
\centering
\begin{tabular}
[c]{|l|l|}\hline
\textbf{Simulation Parameter} & \textbf{Value}\\\hline
Concurrent projections $q$ & $20$\\
$\varepsilon$  & $0.01$\\
Step size $\mu_n$ & $1\mathcal{M}_n$\\
Sparsification $\alpha$ \cite{Slavakis2008} & $0.01$\\
Projection weights  & Based on CI\\
Kernel width $\sigma^2$ & $0.05$\\
\hline
\end{tabular}
\end{table}
\begin{table}
\caption{Multi-Kernel Simulation Parameters}%
\label{tab:param_MultiKernel}
\centering
\begin{tabular}
[c]{|l|l|}\hline
\textbf{Simulation Parameter} & \textbf{Value}\\\hline
Number of Kernels $M$ & $10$\\
$\varepsilon_{\mathrm{MK}}$  & $0.01$\\
$\lambda_1$ & $0.1$\\
$\lambda_2$ & $0.25$\\
Sparsification $\delta$ \cite{Masahiro13} & $0.9995$\\
Kernel widths $\sigma_m^2$ & \begin{footnotesize}$\{10^{-4},5\cdot 10^{-4},10^{-3},\ldots,0.5,1,5\}$\end{footnotesize}\\
\hline
\end{tabular}
\end{table}

{We would like to emphasize that the scope of this paper is the reconstruction of path-loss maps, assuming suitable measurements
are available. 
Moreover, our objective is to reconstruct long term averages as required in network planning and optimization, 
for example, by self-organizing network (SON) algorithms. Therefore, we are not concerned 
with variations of the wireless channel on a small time scale.
}

\subsection{Urban Scenario} \label{sec:SimBerlinMap}

{To assess their performance in real-world cellular networks,} we evaluated the algorithms using realistic
collection of data for the city of Berlin. The data was assembled
within the EU project \textsc{MOMENTUM}, and {it} is available in Extensible
Markup Language (XML) at \cite{momentum}.
The data sets, which include
pixel-based mobility models, {have been} created to meet the need for more
realistic network simulations. 

To be more precise, we resorted to the available data set for an area
of $150\times150$ pixels, each of size $50\,\mathrm{m} \times
50\,\mathrm{m}$. This area represents the inner city of Berlin. For
each pixel, there was path-loss data for 187 base stations.  This data
was generated by ray tracing based on real terrain and building
information. 
{Further details on the dataset and on the scenario can be 
found in MOMENTUM deliverables, particularly in D5.3 and
D4.3, which are also available at \cite{momentum}.}
{We point out that although the resolution of $50\,\mathrm{m}$ may appear low, the dataset provides 
{us with} a unique and complete collection of path-loss data for an entire urban area (not only along  particular measurement paths), 
which {is highly valuable, since we are mainly interested in the capabilities of our algorithms to reconstruct path-loss maps.}
Moreover, we  demonstrate in Section \ref{sec:simulations:small_scale} that,  as long as 
a sufficient number of measurements is available, the performance of our methods 
is independent of the specific environment and the spatial resolution of the underlying data.}
{We} only need a single path-loss value for each
location, {so} each pixel is assigned to a single base station with the
lowest path-loss value. This so-called strongest server assignment {(assuming all base stations 
transmit with equal power)}
determines the geometrical shape of cells because each user reports
its measurement only to the assigned base station. Moreover, each base
station estimates only the path-loss of assigned users, which defines
the coverage area of this base station. Path-loss measurements are
assumed to be reported by each user according to a Poisson
distribution.

As far as the mobility model is concerned, it is assumed that users
move on trajectories {with randomly chosen start and end points} along a real street grid. To produce realistic
movement traces, we used the street data from
\textsc{OpenStreetMap} \cite{URLOSM} %
in conjunction with the vehicular mobility simulator \textsc{SUMO}
\cite{SUMO2011}. {The trajectories were calculated using \textit{duarouter}, a routing
tool that is part of the SUMO simulator.} This combination allows us to generate movement
traces with realistic speed limits, traffic lights, intersections, and
other mobile users.  Furthermore, the traces enable us to perform long
simulations over time intervals of several hours with standard
processors. It is emphasized that, since some roads are more frequently
used than others, the distribution of measurements over the area of
interest is not uniform. As a result, the quality of estimation is not
uniform over the area. 

{In the following}, we study the estimation performance of 
the APSM algorithm and the multikernel approach.  The simulation
parameters are given in Table \ref{tab:param_General}.
\begin{table}
\caption{Simulation Parameters for Urban Scenario}%
\label{tab:param_General}
\centering
\begin{tabular}
[c]{|l|l|}\hline
\textbf{Simulation Parameter} & \textbf{Value}\\\hline
Number of users & $750$, $1500$\\
Simulation duration [s]  & $5000$\\
Number of BS  & $187$, $20$\\
Size of playground & $7500$\,m $\times$ $7500$\,m \\
Size of pixel & $50$\,m $\times$ $50$\,m \\
Measurement frequency $\lambda$ & $0.1$ \\
Number of simulation runs & $10$\\
\hline
\end{tabular}
\end{table}

\begin{figure}
\begin{minipage}{1\linewidth}
\centering
\subfigure[True path-loss map.]{\label{fig:plMaps_org}%
\includegraphics[width=0.49\linewidth ]{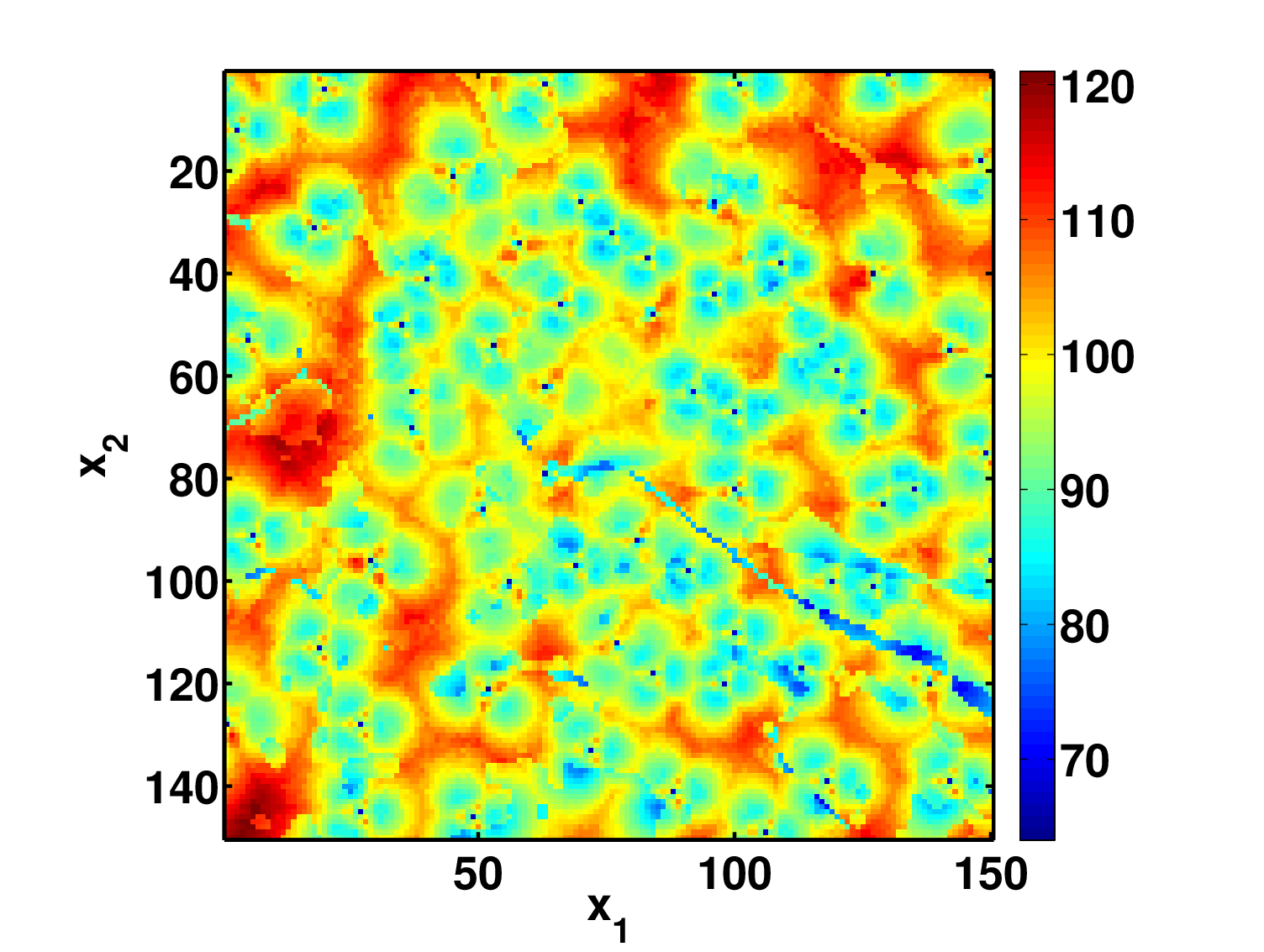}}  
\subfigure[Estimated path-loss map.]{\label{fig:plMaps_apsm}%
\includegraphics[width=0.49\linewidth ]{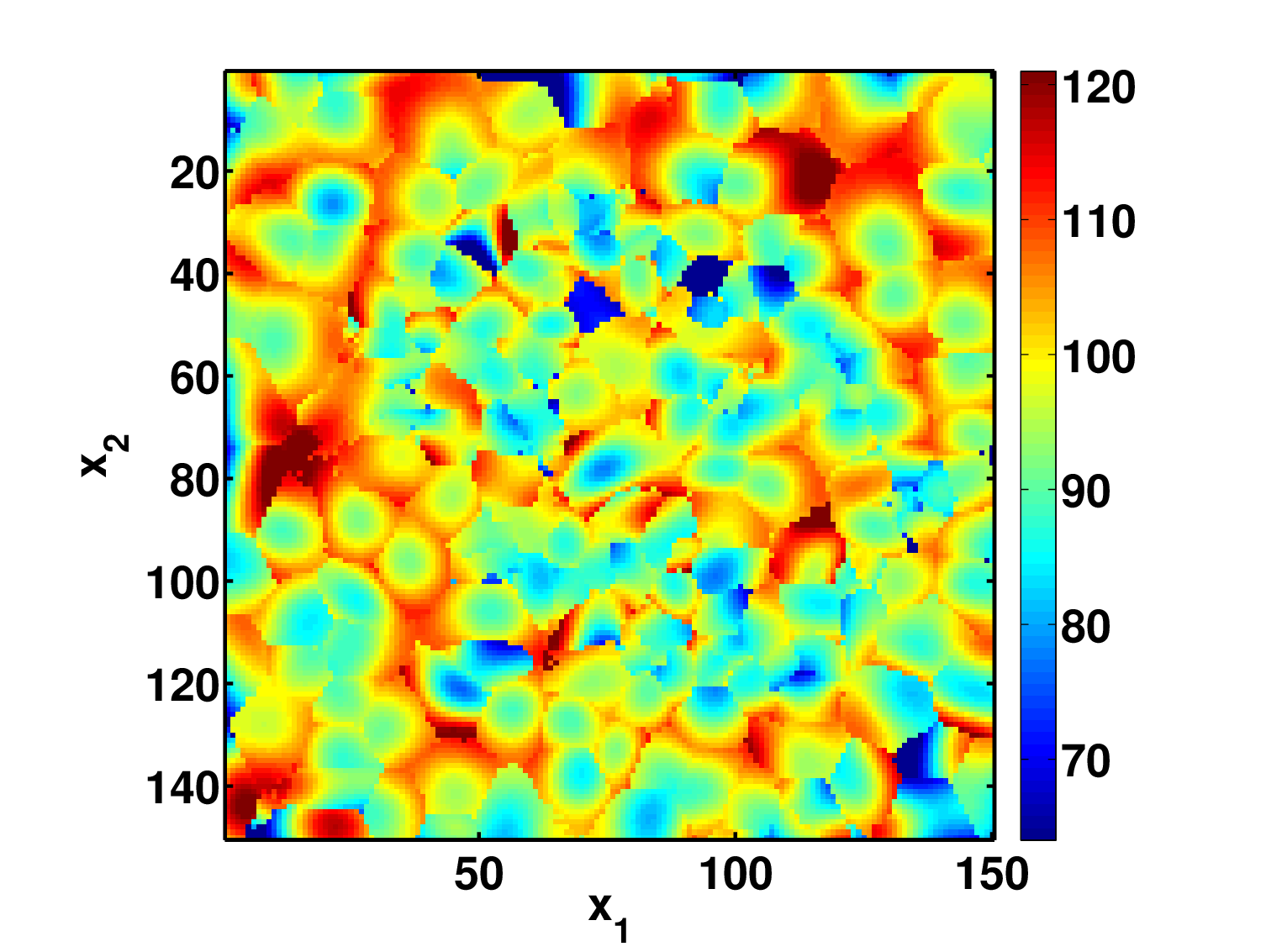}}  
\caption{Visualization of original and snapshot of online-learned path-loss map in urban scenario.}
\label{fig:plMaps}
\end{minipage}
\end{figure}

\begin{figure}[b]
  \centering
  \subfigure[APSM based prediction]{
  \includegraphics[width=0.48\linewidth]{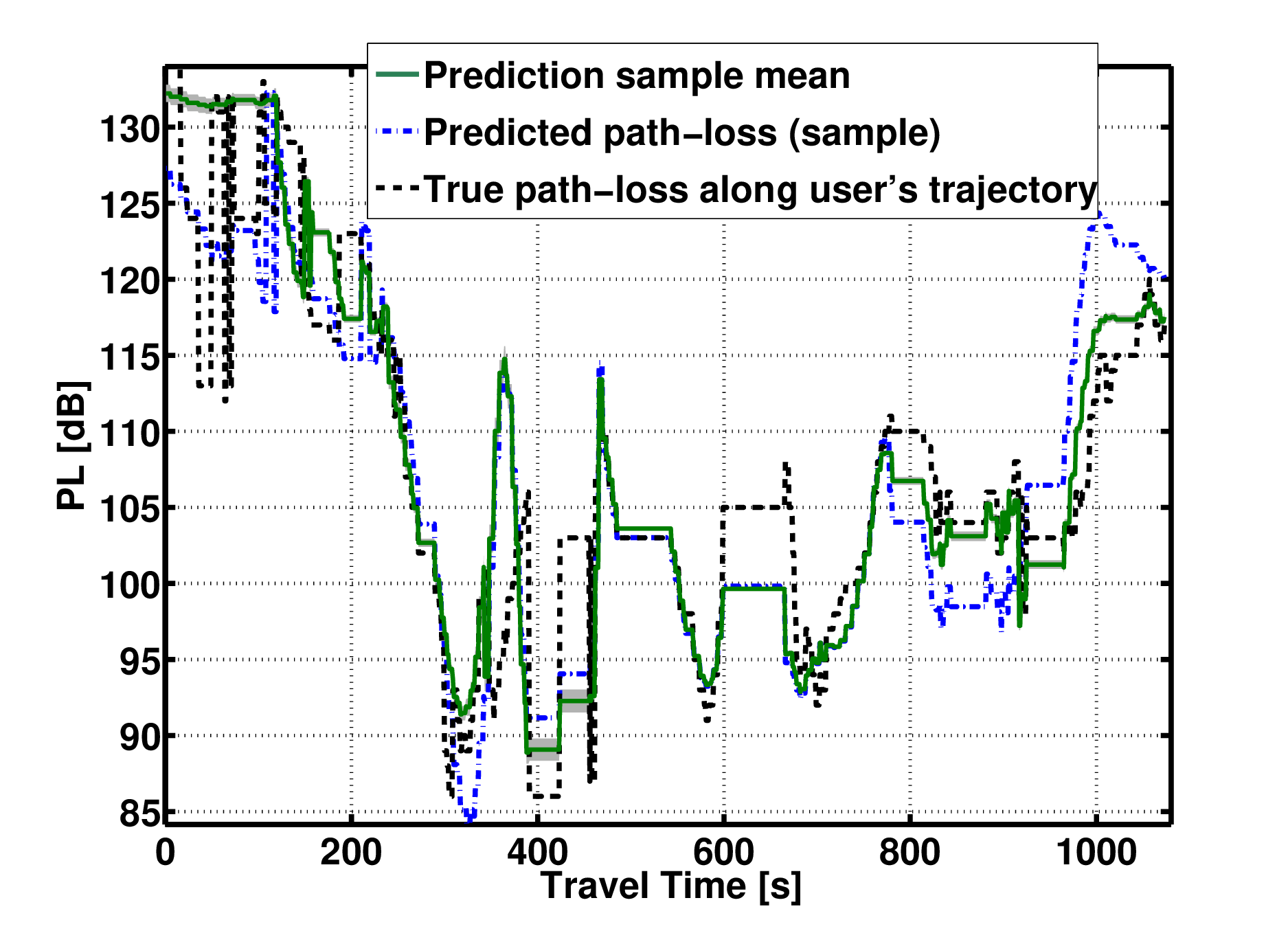}}
  \subfigure[Multikernel-based prediction]{
  \includegraphics[width=0.48\linewidth]{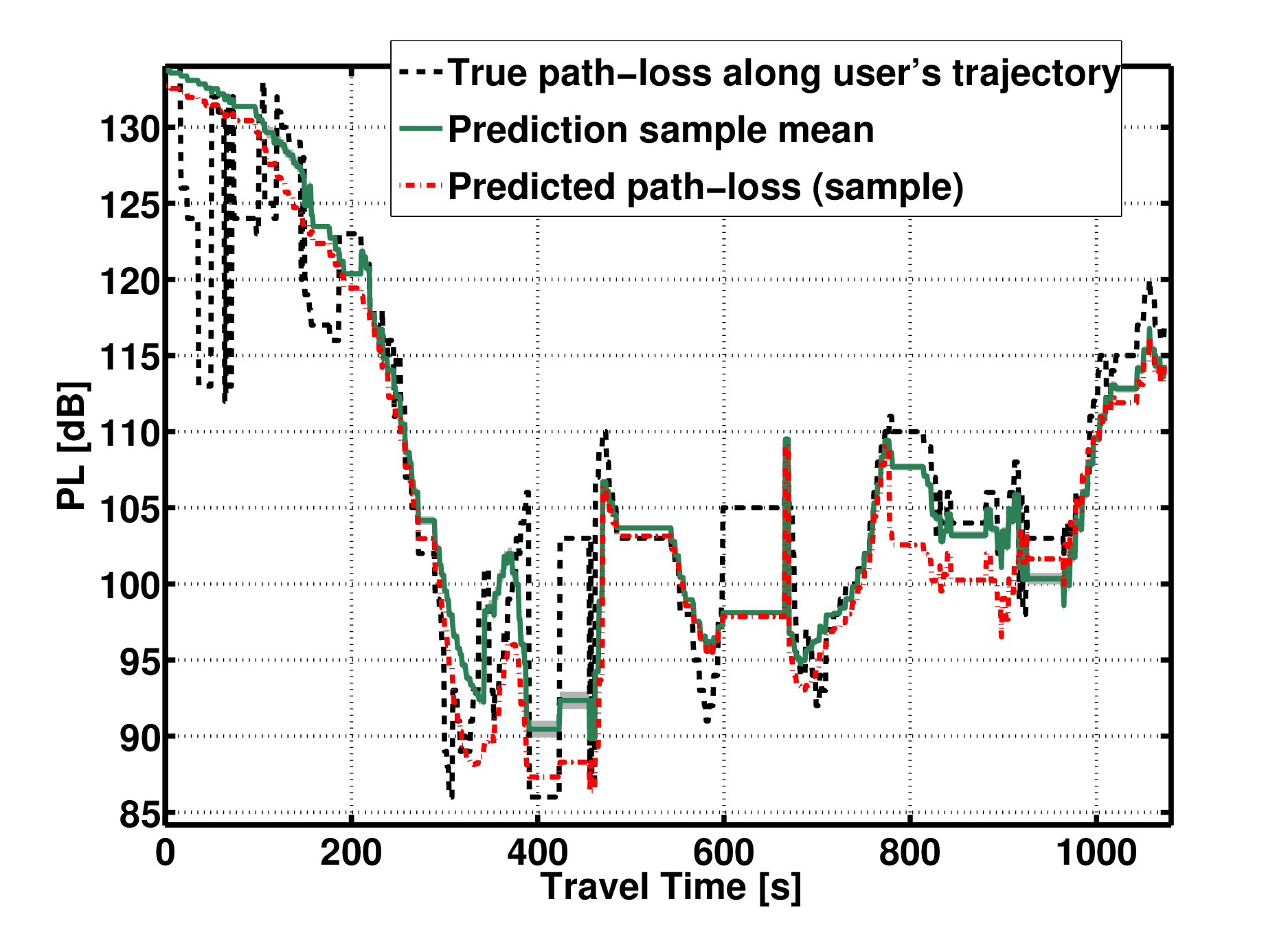}}
  \caption{Comparison of true path-loss experienced by UOI and path-loss predicted by kernel-based methods in urban scenario.}
  \label{fig:userpathloss}
\end{figure}

Figure \ref{fig:plMaps} provides a first qualitative impression of the
estimation capabilities of the proposed algorithms.  We compare the
original path-loss matrix $\bm{H}$ (Figure \ref{fig:plMaps_org}) to
the estimated path-loss matrix $\tilde{\bm{H}}$ (Figure
\ref{fig:plMaps_apsm}) produced by the APSM algorithm after $5000$\,s
of simulated time. Although each base station only learns the
path-losses in its respective area of coverage, the figure shows the
complete map for the sake of illustration.
A visual comparison of the both figures reveals significant
similarities between the estimated and the original path-loss maps. 
The path-loss maps estimated by the base stations can now be used to provide predictions of path-loss
conditions to particular users in the system. 
To confirm the first impression of good performance by Figure \ref{fig:plMaps}, we examine the
estimation results along a route {generated by the SUMO software} between two
arbitrary locations. 
{Figure \ref{fig:userpathloss} illustrates the path-loss prediction for an exemplary route. 
While the dashed-dotted blue and red lines represent a particular sample of the predicted path-loss, obtained using APSM and the multikernel algorithm, respectively, the solid green 
lines represent the mean of the predicted path-loss traces (calculated from 100 runs of the simulation).
In addition, we depict twice the standard error of the mean as shaded area around the mean, which is calculated
for each time {instant}.}
Figure \ref{fig:userpathloss} compares the true
path-loss evolution along the route with the predicted path-loss
evolutions for the two algorithms.
{Note that} the estimated
path-loss values closely match the true ones.  
Furthermore, the
estimation results are expected to improve for routes along frequently
used roads {owing} to {the large} number of {available} measurements.
{For the sake of completeness, we additionally computed the MSE 
to complement Figure \ref{fig:userpathloss}. 
By taking into account only the pixels that belong to the 
specific route, we obtain a (normalized) MSE (as defined in (\ref{eq:definition_MSE}) of $0.0037$ for the APSM algorithm
and  $0.0027$ for the multikernel algorithm.}

\begin{figure}
  \centering
  \includegraphics[width=0.6\linewidth ]{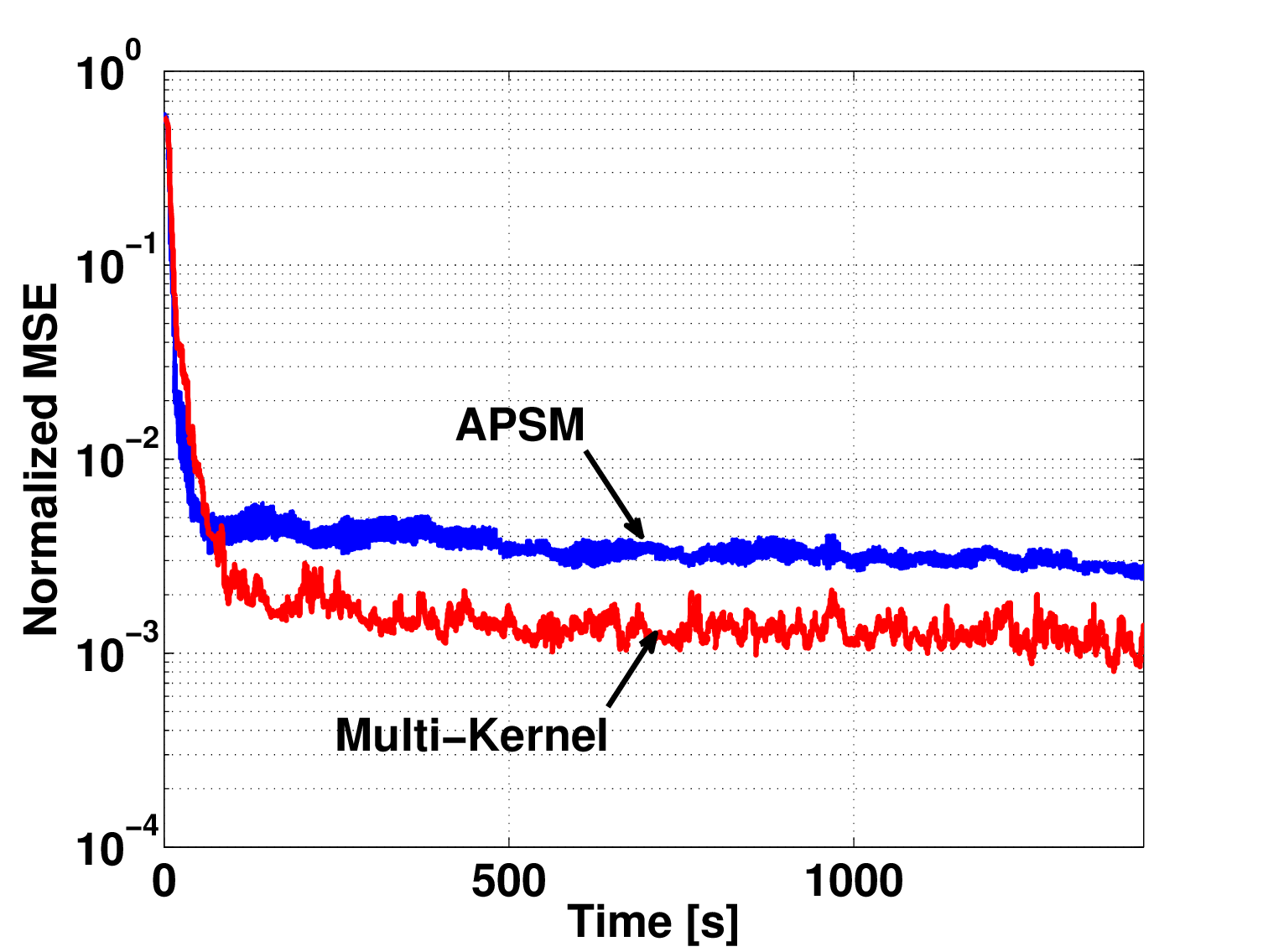}  
  \caption{Comparison of the MSE performances in urban scenario.}
  \label{fig:wholeMSE_BerlinMap}
\end{figure}

Now we turn our attention to the question of how much time the
algorithms require for sufficiently accurate estimation.  To this end, we
studied the evolution of the MSE over time. One result of this study
is presented in Figure \ref{fig:wholeMSE_BerlinMap}, where we observe
a fast decrease of the MSE for both algorithms. The multikernel
algorithm, however, outperforms APSM {in terms of the MSE, 
with only a negligible loss in convergence speed.}

\begin{figure}
  \centering \subfigure[Sensitiveness against errors in reported
  locations. The relative error is defined to be the maximum possible
  error in the measured coordinates (in meter) divided by the total dimension of the
  area ($7500\,\mathrm{m}$).]{\label{fig:mse_over_locerror}%
    \includegraphics[width=0.6\linewidth ]{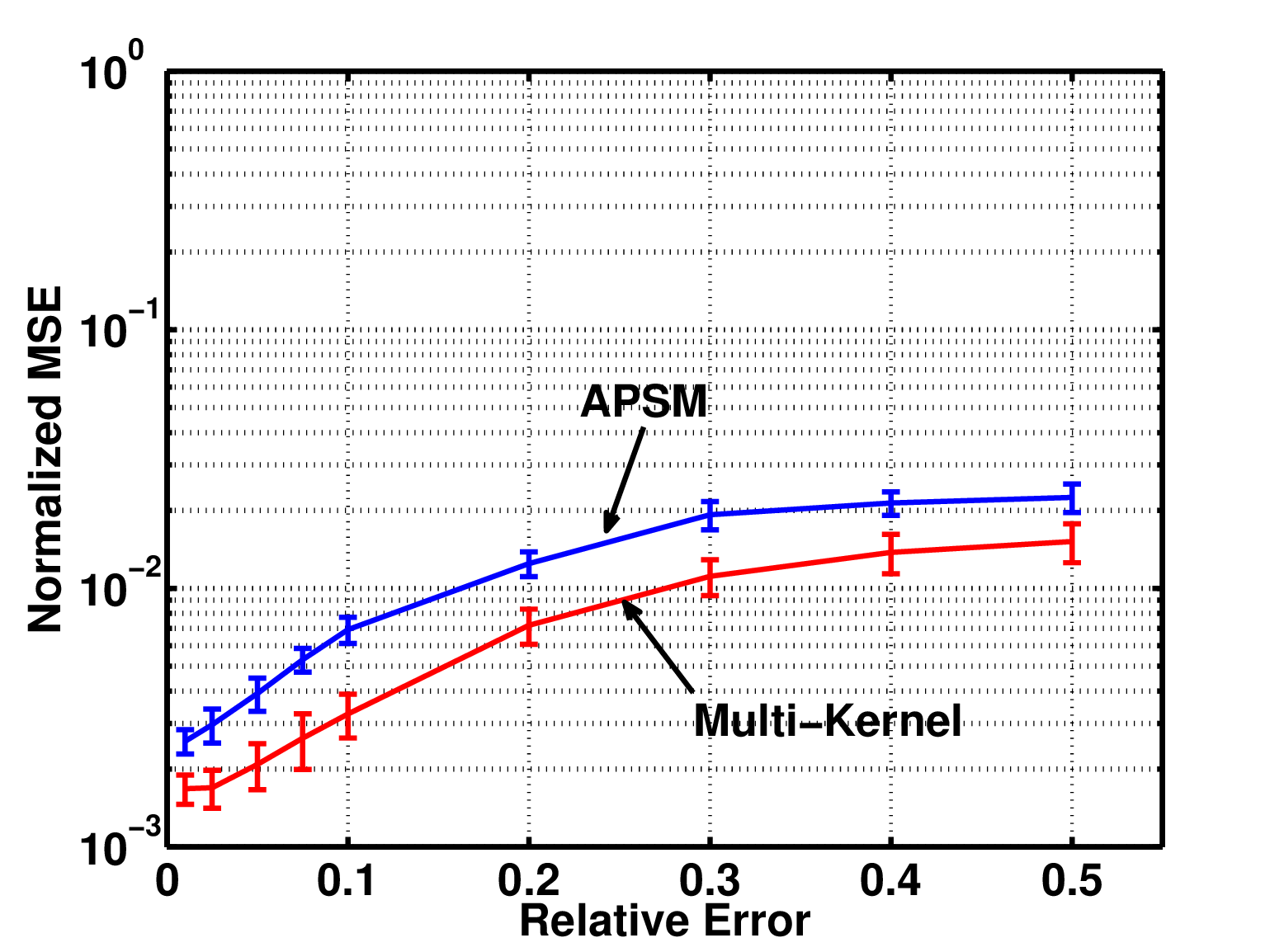} }
  \subfigure[Sensitiveness against errors in path-loss
  measurements. The relative error is the ratio of the maximum
  possible error per path-loss measurement to the average path-loss value
  (averaged over the path-loss values of all pixels).{Thereby a relative error of 
  $0.1$ corresponds to a maximum path-loss error of (roughly) $10\,\mathrm{dB}$.}]{\label{fig:mse_over_measerror}%
  \includegraphics[width=0.6\linewidth ]{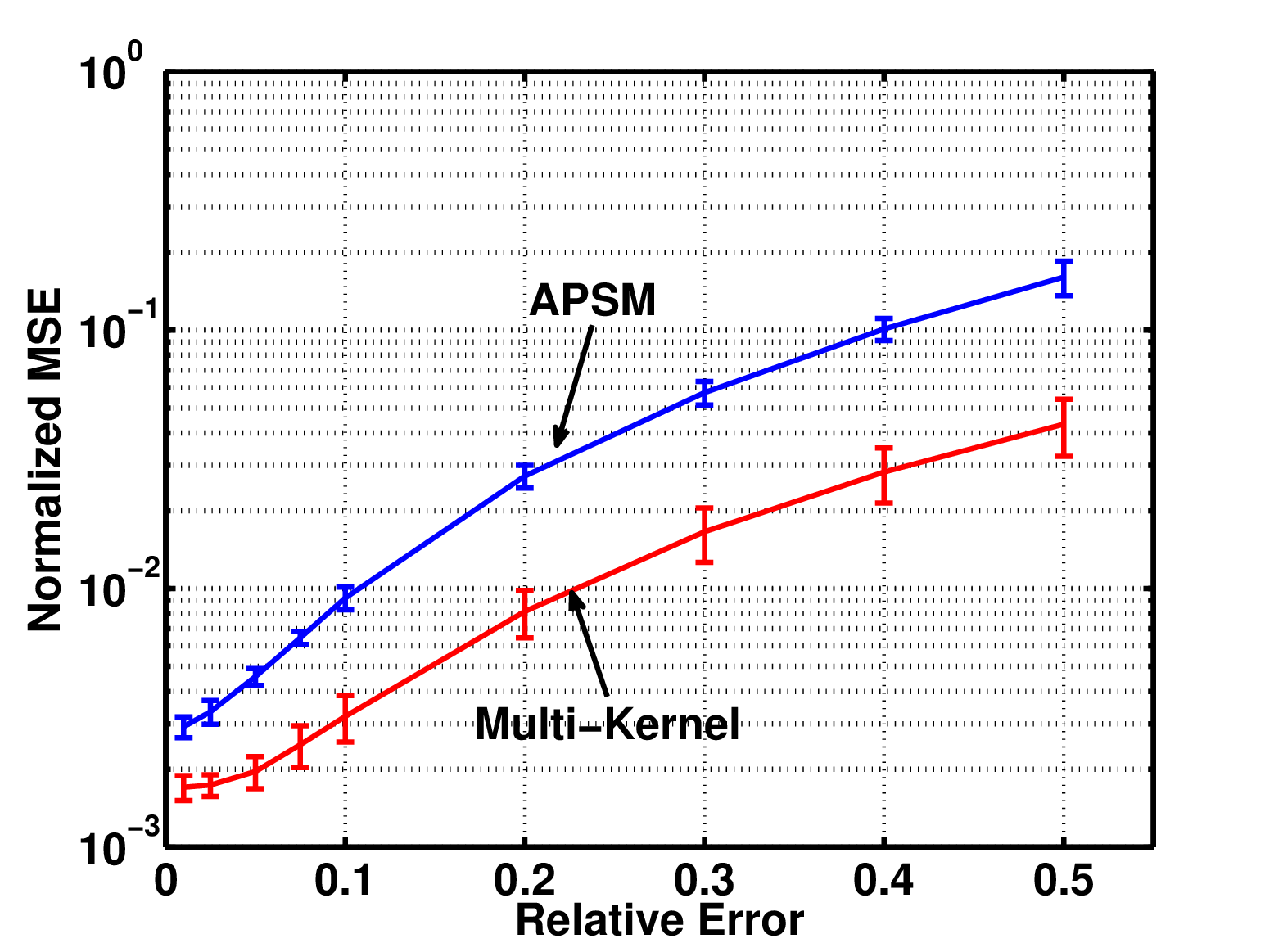}}
\caption{Impact of errors on the MSE performance in urban
  scenario. Error bars represent 95\% confidence intervals under the
  assumption that the sample average of the MSE after $5000\,\mathrm{s}$ is normally distributed.}
  \label{fig:mse_over_error}
\end{figure}

As described in Section \ref{sec:Problem}, the measured value and the
reported location can be erroneous. Therefore, it is essential for the
algorithms to be robust against measurement errors.  Figures
\ref{fig:mse_over_locerror} and \ref{fig:mse_over_measerror} depict
the impact of incorrect location reports and erroneous path-loss 
measurements on the MSE, respectively. {This errors are chosen uniformly at random, with 
a variable maximum absolute value that is investigated in Figure \ref{fig:mse_over_error}.} {Note} that a location error of
less than $5\%$ means an offset in each reported coordinate of at most
$375$\,m (i.e., up to $8$ pixels). Such large errors are usually not
encountered in modern GPS devices \cite{CSR_GPS13}.

From these simulations, we can conclude that both algorithms are
robust to the relevant range of errors but they are more sensitive to
path-loss measurement errors than to inaccurate location reports. In 
this particular scenario, we can also say that the multikernel algorithm is more robust than
the APSM algorithm. Finally, we show that the multikernel algorithm
achieves an accuracy similar to that of the APSM algorithm using a
drastically smaller dictionary.
\begin{figure}[t]
\centering
\includegraphics[width=0.6\linewidth ]{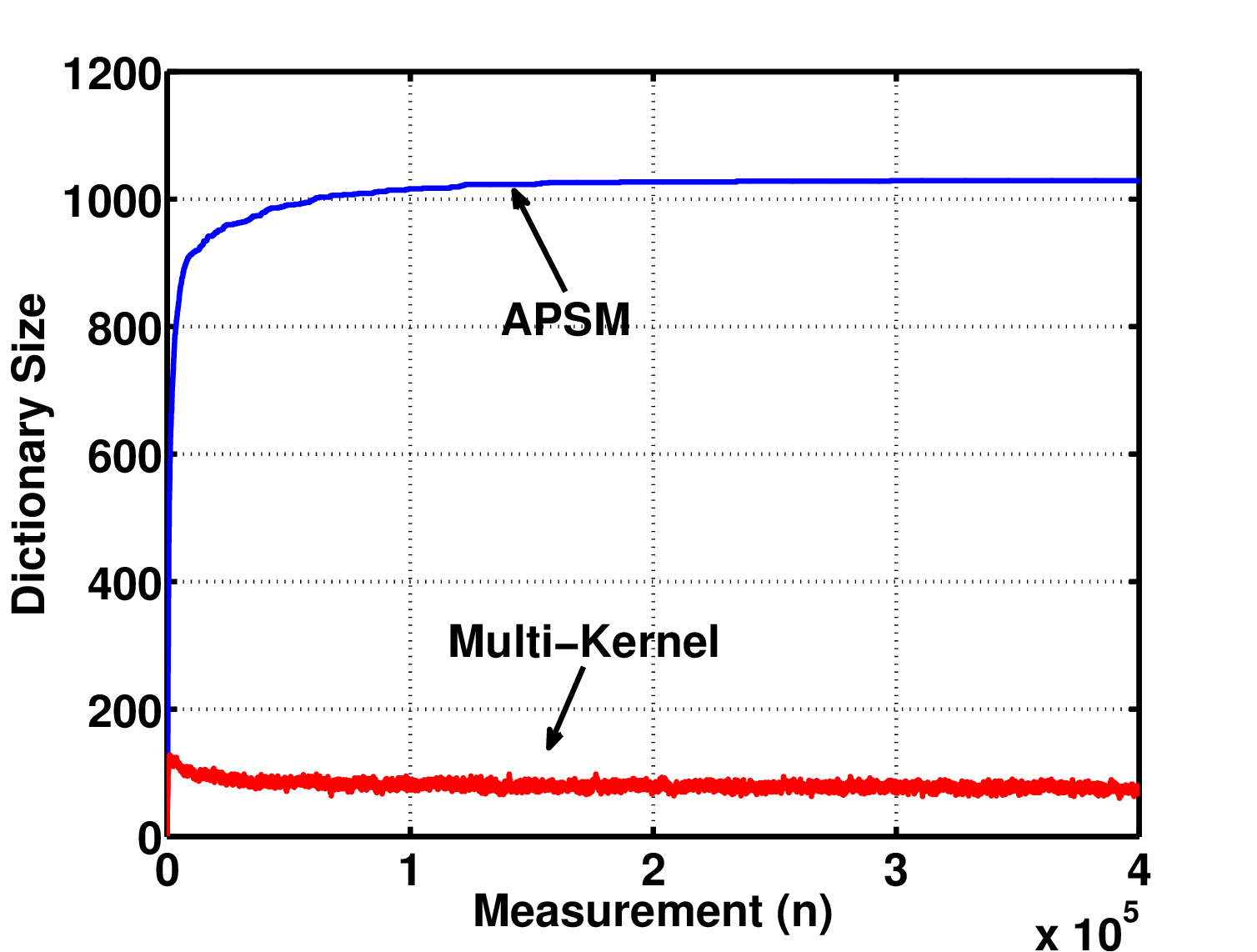} %
\caption{Comparison of dictionary sizes using APSM and multikernel approach with iterative weighting.}
\label{fig:sparse_MM}
\end{figure}
This can be concluded from Figure \ref{fig:sparse_MM}. For an
exemplary simulation run, it shows that the evolution of the
dictionary sizes over time for the APSM algorithm and the multikernel
algorithm. We observe that the dictionary size of the multikernel
algorithm is decreasing with the time, which is due to the block
sparsity enforced by the second term in the cost function given by
(\ref{eq:CF-MK}).

{The parameters of the algorithms have been tailored to
the scenario at hand to obtain the best possible estimation
results. This, however, raises the question of how sensitive the
estimation performance is to parameter changes. To provide insight
into this problem, we performed experiments for selected parameters.
As far as the APSM algorithm is concerned, the parameters are the
tolerable error level $\varepsilon$ in
(\ref{eq:consistencyConditionAPSM}), the step size $\mu$ of the update
in (\ref{eq:UpdateAPSM}), the measure of linear independency $\alpha$ (cf. \cite[Section 5.3]{Slavakis2008}), 
which is used for sparsification, and the kernel width $\sigma$ of the
Gaussian kernel in (\ref{eq:Gaussian}). In case of the multikernel
algorithm, we study the impact of the error level $\varepsilon_{\mathrm{MK}}$ in
(\ref{eq:HyperslabMK}), the step size $\eta$, and the sparsification 
criterion $\delta$ (defined in \cite[Equation 9]{Masahiro13}).  Following the principle of $2^k$ factorial design
\cite{law00}, we chose for each parameter a very low value, indicated
by the subscript $l$, and a very high value, indicated by the
subscript $h$ (the particular parameter values are given in Tables \ref{tab:param_factorial_APSM} and \ref{tab:param_factorial_MK}).
\begin{table}
\caption{Parameter choice in Figure  \ref{fig:paramChoice_APSM}.}%
\label{tab:param_factorial_APSM}
\centering
\begin{tabular}
[c]{|l|c|c|c|c|c|c|c|c|}\hline
\textbf{Parameter} & $\epsilon_{\footnotesize{l,\text{APSM}}}$ &$\epsilon_{h,\text{APSM}}$ & $\mu_{l}$& $\mu_{h}$& $\alpha_{l}$& $\alpha_{h}$& $\sigma_{l}$&$\sigma_{h}$ \\ \hline
\textbf{Value} & $0.0001$ & $0.9$ & $0.1$ & $1.99$ & $0.001$ & $0.9$ & $0.0001$ & $1$ \\
\hline
\end{tabular}
\end{table}
\begin{table}
\caption{Parameter choice in Figure  \ref{fig:paramChoice_MK}.}%
\label{tab:param_factorial_MK}
\centering
\begin{tabular}
[c]{|l|c|c|c|c|c|c|}\hline
\textbf{Parameter} & $\epsilon_{l,\text{MK}}$ &$\epsilon_{h,\text{MK}}$  & $\eta_{l}$&$\eta_{h}$ & $\delta_{l}$&$\delta_{h}$\\ \hline
\textbf{Value} & $0.0001$ & $0.9$ & $0.01$ & $1.99$ & $0.5$ & $0.99999$\\
\hline
\end{tabular}
\end{table}
Then the algorithms were evaluated with all the
parameter choices and the resulting estimation accuracy was compared
to accuracy of the default choice. \\
The results are summarized in Figure \ref{fig:paramChoice}. In
particular, Figures \ref{fig:paramChoice_APSM} and
\ref{fig:paramChoice_MK} show the influence of important design
parameters on the APSM algorithm and the multikernel algorithm,
respectively.
\begin{figure}
\centering
\subfigure[APSM]{\label{fig:paramChoice_APSM}%
\includegraphics[width=0.49\linewidth ]{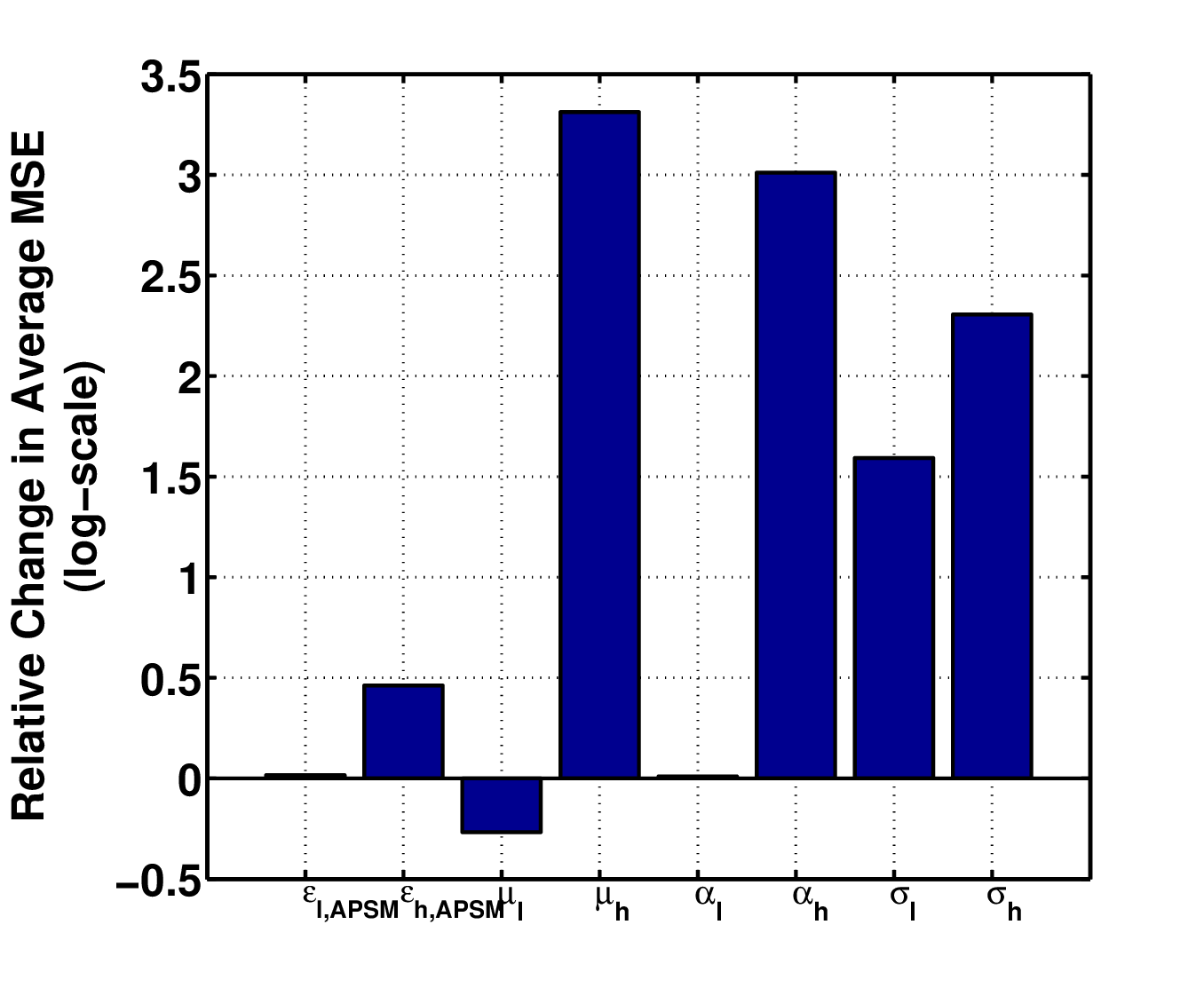}}  
\subfigure[Multi-kernel algorithm]{\label{fig:paramChoice_MK}%
\includegraphics[width=0.49\linewidth]{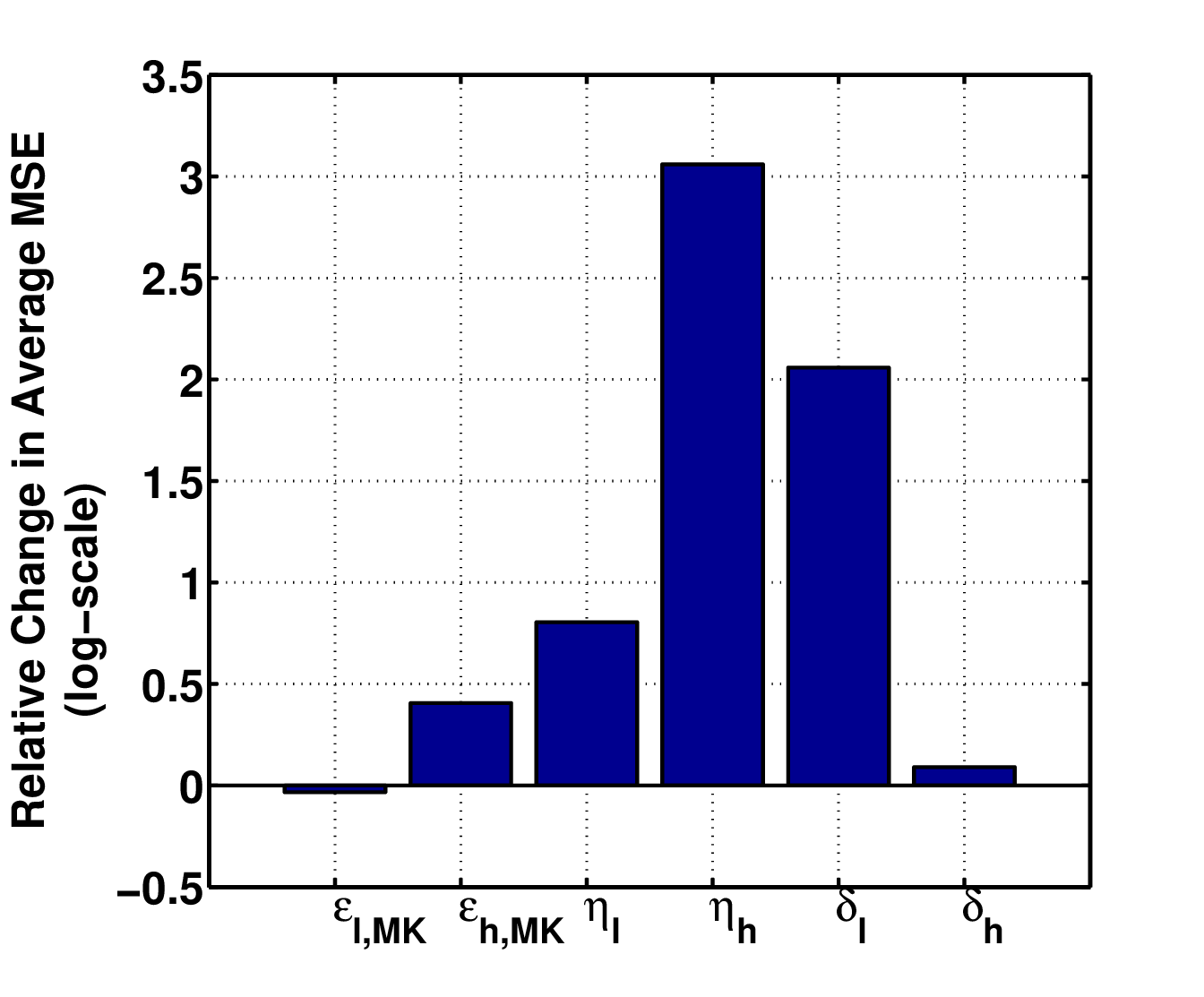}}  
\caption{Effect of the algorithm parameters on MSE in urban setting.} 
\label{fig:paramChoice}
\end{figure}
From these figures, we can conclude that the MSE can in general be
improved by choosing a very small $\mu$ ({cf.} Figure \ref{fig:paramChoice_APSM}). 
{However, these}  performance 
gains  are in general achieved at the cost of a worse convergence
speed, which is ignored in Figure \ref{fig:paramChoice}.} {This observation is a 
manifestation of the known tradeoff between convergence speed versus steady-state performance
of adaptive filters.}

\subsection{Campus network scenario} \label{sec:simulations:small_scale} 
{To demonstrate the capabilities of the proposed methods in a different setting, 
we now use the \textit{cu/cu\_wart} dataset available 
in the CRAWDAD project \cite{cu-cu_wart-2011-10-24}. %
The dataset provides traces of {received signal strength (RSS)} measurements collected in a 802.11 network at the University of 
Colorado Wide-Area Radio Testbed, which
spans an area of $1.8 \times 1.4$ kilometers. 
These traces are measured at a high spatial resolution. 
In fact, over 80\% of the measurements were taken less than 3 meters apart.
{Further details on the data, in particular on the measurement procedure, 
can be found at \cite{cu-cu_wart-2011-10-24}.}
We used 70\% of the data points (chosen uniformly at random) for training and the remaining 30\% of the data points for testing and performance evaluation. 
}
{This scenario has available less data than the previous urban scenario. This gives us the opportunity to compare the performance of our 
low-complexity online methods with (more complex) state-of-the-art batch-processing methods. 
{With reduced data, numerical issues due to finite precision do not arise.} 
In particular, we choose a popular method from geostatistics
for comparison; namely, \textit{ordinary kriging} \cite{phillips2012practical}.}
For example, in \cite{EstevezKriging}, this method was applied to the estimation of radio maps (in particular, SNR maps).

\begin{figure}
  \centering
  \includegraphics[width=0.6\linewidth ]{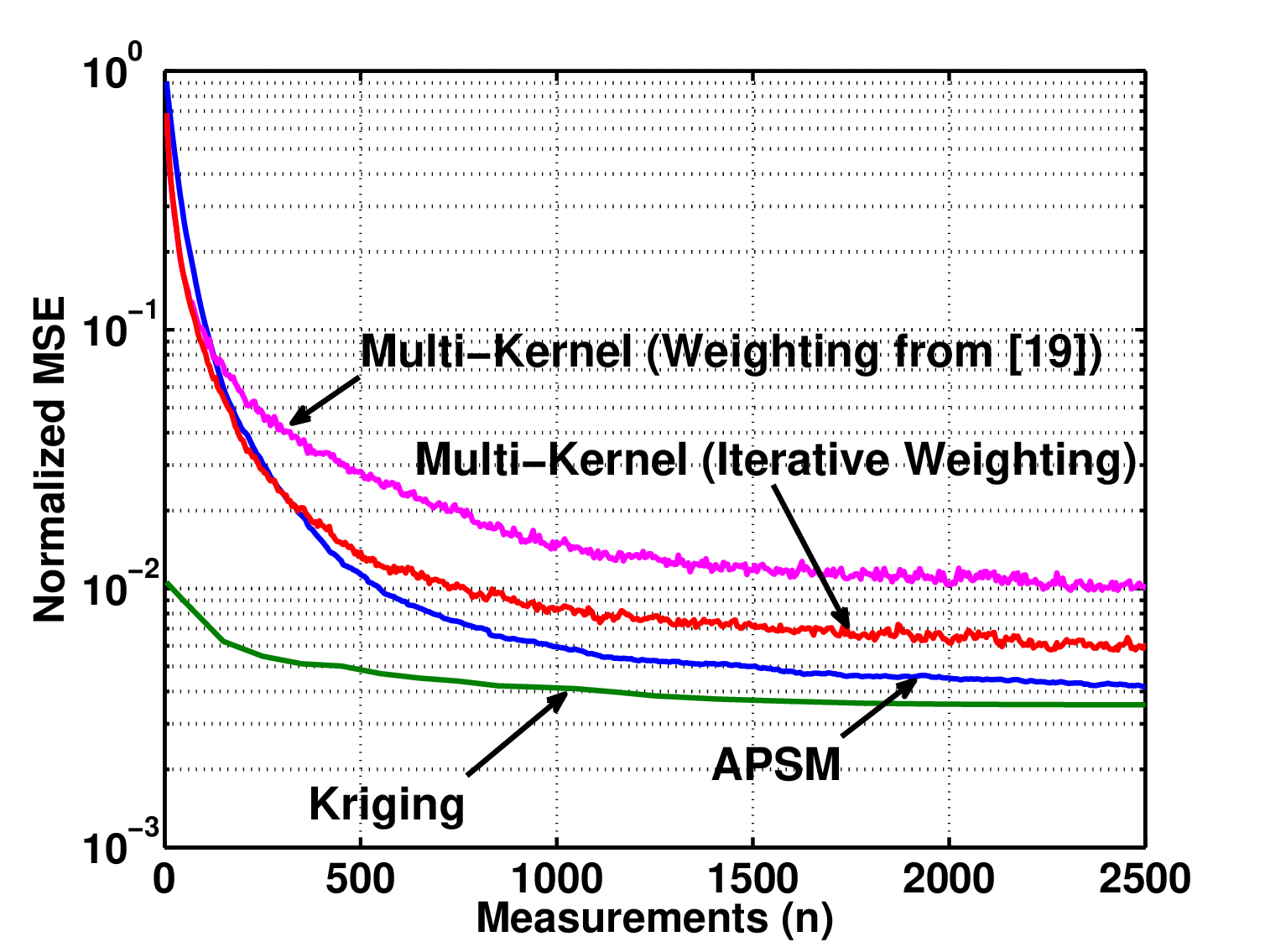}  
  \caption{Comparison of kernel-based algorithms' performance in CRAWDAD scenario. As a baseline, a curve based on a popular but non-realtime geostatistical method (kriging)
  is shown.}
  \label{fig:ResultCRAWDAD}
\end{figure}

{
Figure \ref{fig:ResultCRAWDAD} illustrates the decay of the MSE
for the APSM algorithm and the multikernel algorithm as the number of
measurements increases. {The} MSE is calculated using only the test data points, 
which were not used in the learning process.  
We note that both algorithms behave as desired; 
the MSE decreases fast with the number of measurements. (Our methods perform only one iteration per measurement due to their online nature, 
{but} we may re-use training data points multiple times. Therefore, the number of measurements in Figure \ref{fig:ResultCRAWDAD} is larger then the number of available
training data points.) 
The multikernel approach seems to be outperformed by the APSM algorithm 
with respect to the convergence speed in this scenario. 
As {shown} in Section
\ref{sec:SimBerlinMap}, this {result does not seem to} hold when the underlying
data is more irregular.
{We see two reasons for this changing behavior. Both are related to the fact that the 
data of the Campus scenario is more regular
than that of the Berlin scenario. The path-loss does not have sudden spatial changes. 
In this case, it is easier to tailor the parameters of the APSM method, and, in particular, the 
kernel width to the data at hand.
In fact, with highly irregular data, a very narrow kernel is needed by the APSM to achieve good 
performance, which is detrimental 
to the convergence speed. In contrast, the multi-kernel algorithm has the advantage of  offering 
multiple different kernels to adapt
automatically to the data, without manual tuning of parameters.}
}

{Not only is the estimation quality important, but also the complexity,
which grows fast with the amount of data and the number of
kernels. Therefore, in order to reduce the complexity, we proposed in
Section \ref{sec:algorithm_sparsity} an iterative weighting scheme
that exploits sparsity of the multikernel algorithm. In addition to the comparison to APSM,
Figure  \ref{fig:ResultCRAWDAD} indicates that our iterative weighting approach 
significantly
outperforms that of \cite{Masahiro13} with respect to the convergence speed.}

{So far, we have only compared the two proposed kernel-based online learning schemes with each other. However, it is also of interest 
{to compare the proposed methods} 
with state-of-the-art approaches in the literature. 
Therefore, Figure \ref{fig:ResultCRAWDAD} also includes the MSE performance of %
{the kriging} approach.
{We highlight} that since kriging is essentially an offline scheme where all measurements have to be available beforehand, {so} a 
comparison is difficult and not entirely fair. 
In order to make a direct comparison possible, we apply the kriging technique at certain time instances with all (at those points 
in time) available measurements. Despite providing highly precise results given the amount of available measurements (see Figure \ref{fig:ResultCRAWDAD}), 
this method has a serious drawback, which is inherent to all batch methods: 
the computational complexity does not scale, {because} it grows fast with the number of data points available. 
In fact, all calculations have to be carried out from scratch as soon as a new measurement arrives. 
We apply an \textit{ordinary kriging estimator} with a Gaussian semivariogram model. The most important simulation parameters can also be found in Table \ref{tab:param_General_Artificial}. 
The reader is referred to \cite{phillips2012practical,EstevezKriging} for further details. \\
In Figure \ref{fig:ResultCRAWDAD}, we can observe that the kriging scheme shows very good results even with few measurements, but from a certain 
point in time, more measurements do not significantly improve the quality of estimation. Moreover, we observe that our kernel-based schemes, 
after a sufficient number of measurements (i.e., iterations), can achieve almost the same level of accuracy, but at a much lower complexity and in an online
fashion. To be more precise, we note that the kriging method needs to solve a dense system of linear equations in every iteration. This has 
a complexity of $O(n^3)$ (with $n$ being the number of available samples). In addition, 
to obtain the results in Figure \ref{fig:ResultCRAWDAD} a parametric semivariogram model needs to be fitted to the sample data (see \cite{EstevezKriging} for details) which is a (non-convex)
problem of considerable complexity.  
\begin{table}
\caption{Simulation parameters for the campus network scenario}%
\label{tab:param_General_Artificial}
\centering
\begin{tabular}
[c]{|l|l|}\hline
\textbf{Simulation Parameter} & \textbf{Value}\\\hline
Number of unique data points  & $1274$ \\
Number of simulation runs & $100$ \\
Amount of training data & $70\%$ of data points \\
Amount of test/validation data & $30\%$ of data points \\
Simulation duration (iterations) & 2500 \\ 
Semivariogram fitting model & Gaussian \\
\hline
\end{tabular}
\end{table}
}

\section{Conclusions} \label{sec:Conclusions} 

Based on the state-of-the-art framework for APSM-based and 
multikernel techniques, we have developed
novel learning algorithms for the problem of reconstructing a
path-loss map for any area covered by a large cellular network. In
addition to path-loss measurements performed by sparsely and
non-uniformly distributed mobile users, the proposed algorithms can
smoothly incorporate side information to provide robust reconstruction
results. Low-complexity is ensured by imposing sparsity %
on kernels and measurement data.  Unlike current solutions based on
measurements, the algorithms can operate in real time and in a
distributed manner. Only to predict the path-loss for locations
outside of the current cell of the user of interest, a data exchange
between adjacent base stations becomes necessary.

As pointed out in the introduction, the path-loss information provided
by the proposed algorithms can be used to reconstruct a complete
coverage map for any area of interest. The accurate knowledge of
coverage maps allows us to design robust schemes for proactive
resource allocation. In particular, if combined with prediction of
mobile users' trajectories, the delay-tolerance of some applications
can be exploited to perform spatial load balancing along an arbitrary
trajectory for a better utilization of scarce wireless resources. The
performance of such schemes can be further enhanced if, in addition to
a path-loss map, the network load is predicted. This is an interesting
and very promising research direction for future work.

\section*{Acknowledgment} 

The authors would like to thank Hans-Peter Mayer from Bell-Labs,
Stuttgart for the inspiring discussions.

\appendices

\bibliographystyle{IEEEtran}
\bibliography{IEEEabrv,prereal}

\end{document}